\documentclass[prb,twocolumn,floatfix]{revtex4-2}

\usepackage{placeins}
\usepackage{latexsym}
\usepackage{graphicx}

\usepackage{subfigure}    

\usepackage{amsmath}
\usepackage{color}
\usepackage{xcolor}
\usepackage{soul}

\usepackage{relsize}
\usepackage{placeins}
\usepackage{amssymb}
\usepackage{amsthm}
\usepackage{amsfonts}
\usepackage{braket}
\usepackage{latexsym}
\usepackage{graphicx}
\usepackage{grffile}
\usepackage{hyperref}
\usepackage{multirow}

\begin{document}

\graphicspath{{fig/}{./}}
\title{Lattice-mismatched and twisted multi-layered materials for efficient solar cells}
\author{Efstratios Manousakis}
%	\email{manousakis@gmail.com}
	\affiliation{Department  of  Physics, Florida  State  University,  Tallahassee,  FL  32306-4350,  USA}
%	\affiliation{Department   of    Physics,  National and Kapodistrian University    of   Athens,
%		Panepistimioupolis, Zografos, 157 84 Athens, Greece}
	%\ead{aryalniraj7@gmail.com} %,manousakis@magnet.fsu.edu}
	%\maketitle
\date{\today} % Date, can be changed to a custom date

\date{\today}
%\section{Introduction}
\begin{abstract}
  We argue that alternating-layer structures 
  of lattice mismatched  or misaligned (twisted) atomically-thin layers
    should be expected to be more
  efficient absorbers of the broad-spectrum of solar radiation
  than the bulk material of each individual layer.
  In such mismatched layer-structures the conduction and valence bands of the bulk material, 
  split into multiple minibands separated by minigaps  confined 
 to a small-size emerging Brillouin zone due to band-folding.
 We extended the Shockley-Queisser approach to  calculate the photovoltaic efficiency for a band
 split into minibands of bandwidth $\Delta E$ and
mini-gaps $\delta G$ to model the case when such structures are used as solar cells.
\iffalse
we show that a significant reduction
 of the energy-loss due to phonon-emission processes and an 
 enchancement of the impact ionization rate, a phenomenon
 which leads to multi-carrier generation, should be expected.
 \fi 
    We find a significant
efficiency enhancement due to impact ionization processes, especially in the limit of small but non-zero $\delta G$, and a dramatic increase when fully concentrated sun-light is used.
\end{abstract}
\maketitle

\section{Introduction}

Conversion of the energy of solar radiation into electrical form
by means of the photovoltaic (PV) effect is widely used in most regions of
our planet wasting approximately 80\% of it into heat.
\iffalse
The problem is not
simply limited to this waste of energy but, in addition, this large amount
of heat produced leads to faster degeneration of the material used for this
process. \fi
A key reason for this low efficiency\cite{10.1063/1.1736034} is 
that the photon energy, from the broad solar radiation-spectrum,
absorbed by a single
electron-hole pair with energy above the semiconductor
band-gap is lost by means into heat. The main mechanisms of this loss is
 electron-phonon scattering and
phonon-emission through which electrons and
holes relax to the conduction band-minimum (CBM) or
valence-band maximum (VBM). If instead, semiconductors with a relatively
large band-gap are chosen in order to avoid the above issue,
most of the energy spectrum of the solar radiation
is cut off and travels through the material without absorption.

There is a growing effort to fabricate structures with higher PV efficiency.
The so-called
multi-junction (MJ) solar-cells have reached efficiency at 40\% levels.
The MJ or tandem solar cell\cite{10.1063/5.0048653} is made by combining a stack of cascaded
multiple pn-junctions with various band gaps in order to catch a
larger photon energy spectrum. It is an expensive approach and these
cells are preferred in space applications.

Various ideas to increase efficiency have been proposed\cite{doi:10.1126/science.ado4308}, such as,
using the role of the interaction of localized electronic states with the
conduction band\cite{PhysRevLett.85.1552,PhysRevLett.82.1221}. Highly
gap-mismatched alloys\cite{PhysRevLett.106.028701}, where a class of
materials forms through alloying of distinctly different
semiconductors, have been shown to yield an increased PV efficiency.
In addition, the family of the so-called intermediate band solar
cells\cite{10.1063/1.4916561}, which is based on the idea
 of forming an impurity band in the semiconductor band
gap, could lead to an increased efficiency\cite{PhysRevLett.78.5014}.

Other structures, such as nanotubes, quantum dots, and two-dimensional
(2D) halide perovskites have been considered as candidates for solar
cells\cite{Iqbal2021} with higher PV efficiency because of multi-carrier generation (MCG)\cite{PhysRevB.44.10945,PhysRevB.42.8947,PhysRevB.51.13281}.
Recently, the impact ionization process,
which leads to MCG, was proposed to be effective in strongly correlated insulators\cite{PhysRevB.82.125109,PhysRevB.90.165142,Manousakis2019}.
More recently, it has been proposed\cite{PhysRevB.107.125151,PhysRevLett.130.076901} that utilizing Berry curvature dipoles
can lead to optical gain.

%\section{Idea}

It is well-known that films of multilayer structures can be grown 
by various techniques, including molecular bean epitaxy, chemical
vapor deposition or transport or by a more recent technique where
entire atomically-thin layers of a crystalline material
can be placed on top of another layer.
Our discussion will be focused on structures $ABAB...$ of alternating  atomically-thin of lattice mismatched layers $A$ and $B$ or of misaligned  layers $A$ and $B$ twisted by opposite angles.
 Multiple recent reports
of successfully realizing twisted layers of materials\cite{zhang2023square,Cao2018,Li2016,Sanchez-Santolino2024,lee2024moire,PhysRevMaterials.8.034001} indicate that fabrication of such multilayers of our choice may not
be a too-far-in-the-future project. 
\iffalse
Therefore, given that this is a theoretical paper, we will include
this family of structures in our discussion.
\fi
In both of these different situations, a periodic structure
with an emerging wavelength significantly larger than the Wigner-Seitz unit
cell of each of the layers, can emerge.
The electronic band-structure of such structures lives in a Brillouin zone (BZ)
which is the result of folding the BZ of each of the individual layers
multiple times. Here, we will demonstrate that such
a band-structure has characteristic properties that 
can be tuned to become more efficient absorbers of the solar radiation spectrum.

\iffalse
Therefore, we believe that the theoretical ideas to be discussed in this paper
are not necessarily out of the realm of practical future engineering. 
\fi
A multilayer of twisted or lattice-mismatched layers
can be used to split a band into minibands
which can be used to increase the photo-voltaic (PV) efficiency. 
\iffalse
 The excited
electron is promoted from the occupied valence mini-bands to an excited
conduction miniband.
\fi
The photoexcited electron-hole pair can decay by
means of multi-phonon emission to the bottom of the miniband to which is
was excited. However, since the bandwidth of each miniband is very small,
the energy converted to
heat will only be of the order of the bandwidth of each miniband, from
where it can only decay by means of recombination or impact ionization.
\iffalse
This is to be contrasted with the case where the band is a full band
not split  where the excited ``hot'' electron
in the case of the full
band will not be considered hot in the case where
there is a mini-band
bottom which will ``hold'' it from decaying any further.
%Assuming that the created minigaps are also small to avoid the possibility
%that the photons go through without being absorbed.
\fi
Namely, since there are no states in these multiple mini-gaps,
the photo-excited electrons or holes, after they reach the bottom of the particular mini-band,  have only two possible decay mechanisms:
impact ionization, i.e., by exciting other electron-hole
pairs (which leads to carrier multiplication) or by means of
electron-hole recombination. Therefore, these materials, have
significantly limited the phonon-emission mechanism and, therefore,
as we demonstrate here, should be expected to perform more efficiently as absorbers of a larger
part of the solar spectrum.

Splitting the bands with small gaps allows those photons with energy inside
these gaps to go through the material without been absorbed.
\iffalse
If there are several pairs of twisted bilayers with different
twist angles, the mini-gaps of one bilayer pair can fall within
the bands of another pair. In such case, theoretically, it could
be managed that there would be no or few energy
windows in the solar spectrum for a photon to go through the stack without
being absorbed.
\fi
However, as we will show, the fraction of the gaped
energy interval to that in which there is band presence is  small.
The presence of such mini-gaps prevents or delays the decay of the electron or hole carrier to the bottom of the original band present in the bulk material
before splitting. Furthermore, the band folding generates complex
conservation and
selection rules for a decay process through phonon emission, which only results
in a significant increase of the decay time-scale.
This fosters a convenient environment for the impact ionization processes
to occur, which lead to
carrier multiplication\cite{PhysRevB.82.125109,PhysRevB.90.165142,Manousakis2019}.

\iffalse
If such a scheme turns out to be too difficult to implement,
using a stack of clockwise and counter-clockwise twisted multilayers on top
of a garden variety semiconductor film should also produce a
more PV efficient device. In this simpler case, the photons
that go through the mini-gaps of the front-end material will be absorbed
(albeit with Shockley-Queisser\cite{10.1063/1.1736034} (SQ) limited efficiency) and will be converted to electrical power.

Assuming a 50-50 split, i.e., 50\% of the energy window is energy gap space
and the other 50\% is energy band space, and in addition the original gap
is a very small-fraction of the solar spectrum of 3.5 eV, say 0.3 eV, then
the close to 50\% of the photon will excite e-h pairs to states within
miniband and if there are many mini-bands so that their average bandwidth
is a very small number of the solar spectrum, the efficiency should be close
to 40-45 \%. If we have to such twisted bilayer with complementary mini-gaps
and minibands, the efficiency could theoretically reach the 90\% range. 
\fi

The paper is organized as follows. In the following section we use tight-binding
models to illustrate how large bands of bulk semiconductors can split into
many mini-bands separated by mini-gaps because of the multi-folding of the
Brillouin zone as a result of superlattice formation due to either lattice-mismatch or
twisting of atomically thin layers. In Sec.~\ref{interactions:bands} we
discuss the effects of interactions on these mini-bands. In Sec.~\ref{interactions:relaxation} we
discuss the role of the various interaction and decay processes on the
relaxation
of the photo-excited electron-hole pairs.
In Sec.~\ref{sec:efficiency} we extend the
calculation of Shockley and Queisser\cite{10.1063/1.1736034}, which was
done for a single band, to calculate the photo-voltaic
efficiency enhancement when 
the semiconductor band is split into many bands (all of which are within the energy domain of the solar radiation) of
small bandwidth  separated by small gaps.
Lastly, in Sec.~\ref{sec:conclusions} we discuss our findings and give our
concluding remarks.

%    \bibliography{refs}

\section{Minibands from large band-folding}
\subsection{A ladder of mismatched chains}
\label{mismatched-chains}

\begin{figure}
  \begin{center}
    \includegraphics[width=3.0 in]{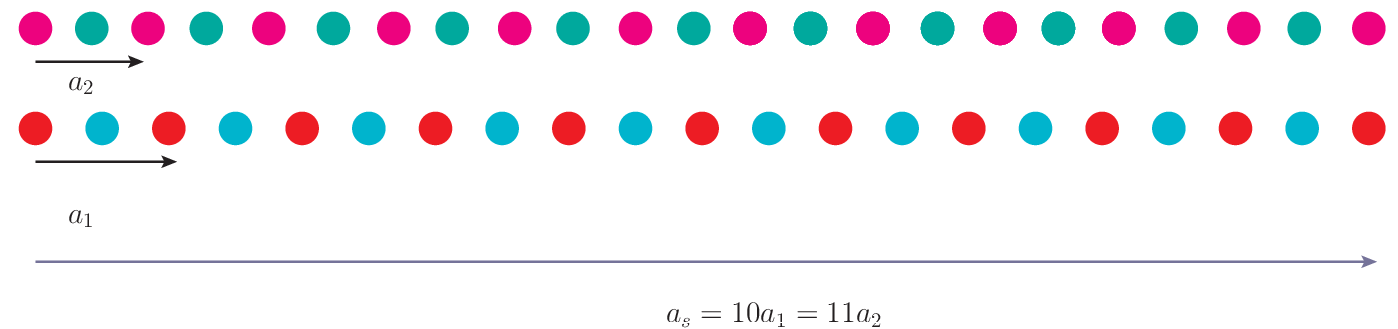}
    \caption{A ladder formed by two chains having a slight lattice mismatch.}
  \label{fig1}
 \end{center}
 \end{figure}

\begin{figure}
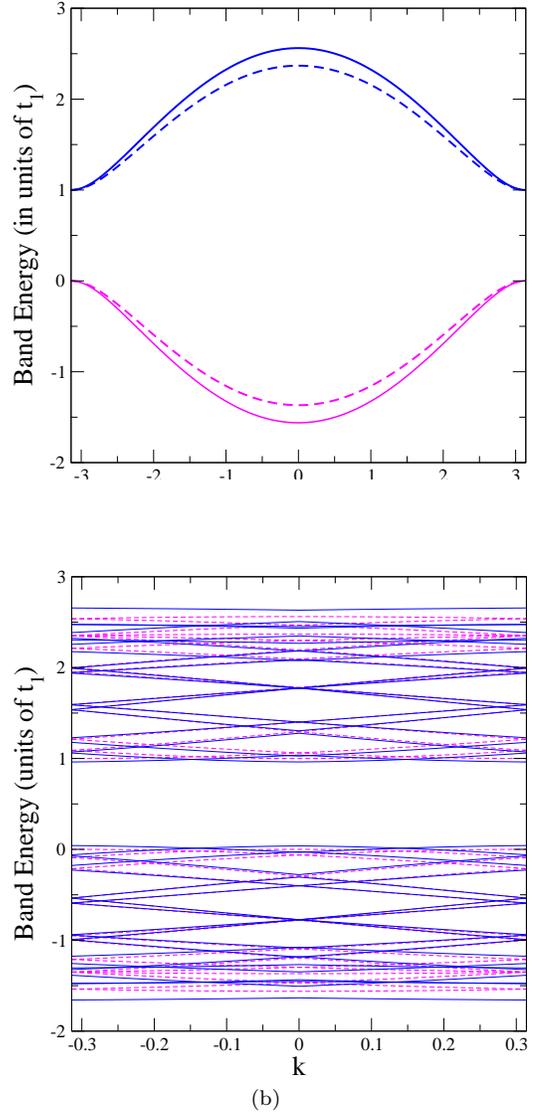

  \begin{center}
            \subfigure[]{
              \includegraphics[scale=0.4]{Fig2a.eps} 
              }
            \subfigure[]{
              \includegraphics[scale=0.4] {Fig2b.eps}
            }
    \caption{(a) The bands of each chain separately. The solid (dashed) lines
    correspond to layer 1 (layer 2) and are plotted versus $k_1 a_1$ ($k_2 a_2$). (b) The band structure  of the combined interacting chains.}
  \label{fig2}
  \end{center}
\end{figure}

As our first example,  we use a simple one dimensional (1D) structure to illustrate the
effect of lattice mismatch on the electronic band-structure.
Two infinite 1D chains are coupled to form the ladder shown in Fig.~\ref{fig1}. Each chain consists of two different atoms ($a$ and $b$) drawn as blue and red for chain 1 (bottom)
and magenta and green for chain 2 (top).  The unit-cell sizes $a_{1,2}$ for each layer are somewhat different, such
that $(M+1) a_2 = M a_1$; we have chosen $M=10$, in such a way that
there is the superperiodic lattice with period $a_s = 10 a_1 = 11 a_2$.
Namely, the ladder shown
in Fig.~\ref{fig1} with 10 unit cells in chain-1 and 11 unit cells in chain-2 repeats itself as the end of the $10^{th}$ unit-cell of chain-1 lines up
with that of end of the $11^{th}$ unit cell of chain-2.
Therefore, the ladder is characterized
by a superperiodic structure, with 20 atomic states in chain 1 and
22 atomic states in chain 2. 
Each chain is described by
a simple tight-binding model  with different on-site energies $e^{(a)}_{1,2}$ and
$e^{(b)}_{1,2}$ for the $a$ and $b$ atoms and 
with a single hopping integral $t_{1,2}$ from atom $a$ to atom $b$ within each chain 1 and 2, which yields the band-structures:
\begin{eqnarray}
  e^{\pm}_{\mu}(k) = {{e^{(a)}_{\mu} + e^{(b)}_{\mu}} \over 2 }
    \pm \sqrt{({{e^{(a)}_{\mu} - e^{(b)}_{\mu}} \over 2 })^2 + 4 t_{\mu}^2
        \cos^2({{k_{\mu} a_{\mu}}\over 2})},
    \end{eqnarray}
where $\mu=1,2$.
These band-structures are illustrated
in Fig.~\ref{fig2}(b) for $t_2/t_1=0.9$ and $e^{(a)}_{1,2}/t_1=1$ and $e^{(b)}_{1,2}=0$.
The band structure of a non-interacting
such ladder is shown by the dotted magenta lines in Fig.~\ref{fig2}(b), which consists of 42 minibands obtained by folding the original
bands (Fig.~\ref{fig2}(a)) into an $M$-fold smaller BZ
(notice the scale of the $k$ axis in Fig.~\ref{fig2}(b) as compared
to that of the bands in Fig~\ref{fig2}(a)). We have introduced constant interaction matrix elements of strength $\delta/t=0.1$ between adjacent bands  to obtain
the bands-structure shown by the blue lines in Fig.~\ref{fig2}(b).
Such interaction can lift the degeneracy between bands at the smaller-BZ
corners creating small gaps. If we allow atomic relaxation, these
gaps are expected to become somewhat wider.

In summary, by having a lattice mismatch between the two 1D layers, we obtained a folded BZ with a number of bands equal to the number of atomic states
in the super-unit-cell obtained as a result of the lattice mismatch.
The band-structure can have a main gap of similar size
as each isolated chain, while  the conduction and valence bands of the coupled chains are split into many mini-bands with mini-gaps.
These characteristics are independent of dimension as demonstrated in two
2D examples, next.

\subsection{Mismatched bi-layers}
\label{mismatched-layers}
\begin{figure}[htb]
  \begin{center}
            \subfigure[]{
              \includegraphics[scale=0.4]{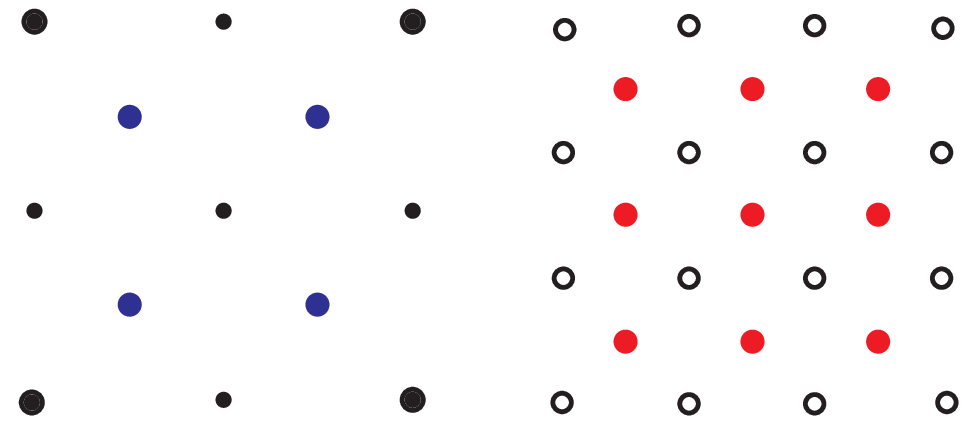}
            }\\
                        \subfigure[]{\hskip -0.2 in
                          \includegraphics[scale=0.24]{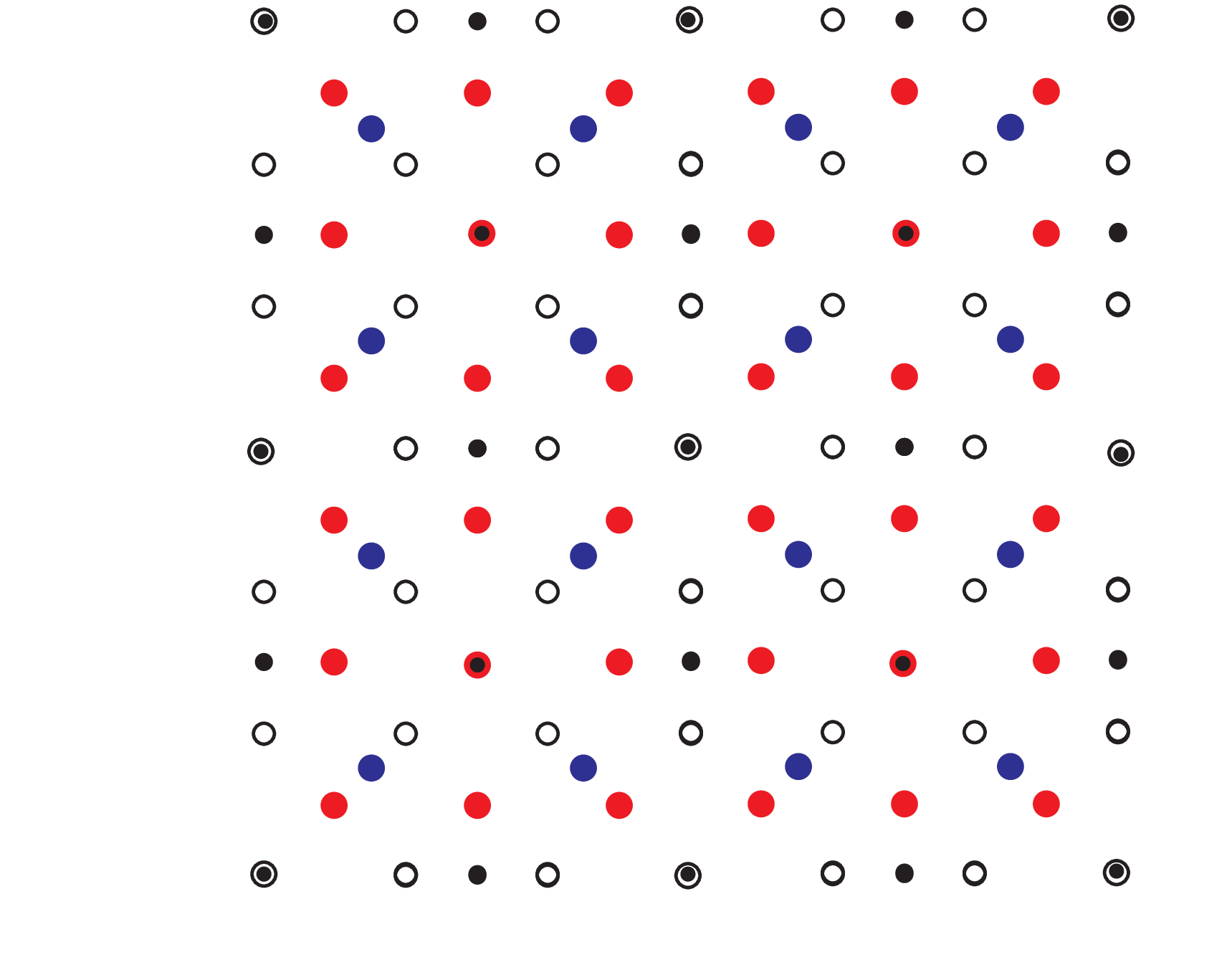}
             }
    \caption{(a) The top two lattices which when super-imposed
      give rise to the unit-cell shown at the bottom (b).}
    \label{fig}
  \end{center}
\end{figure}

\begin{figure}
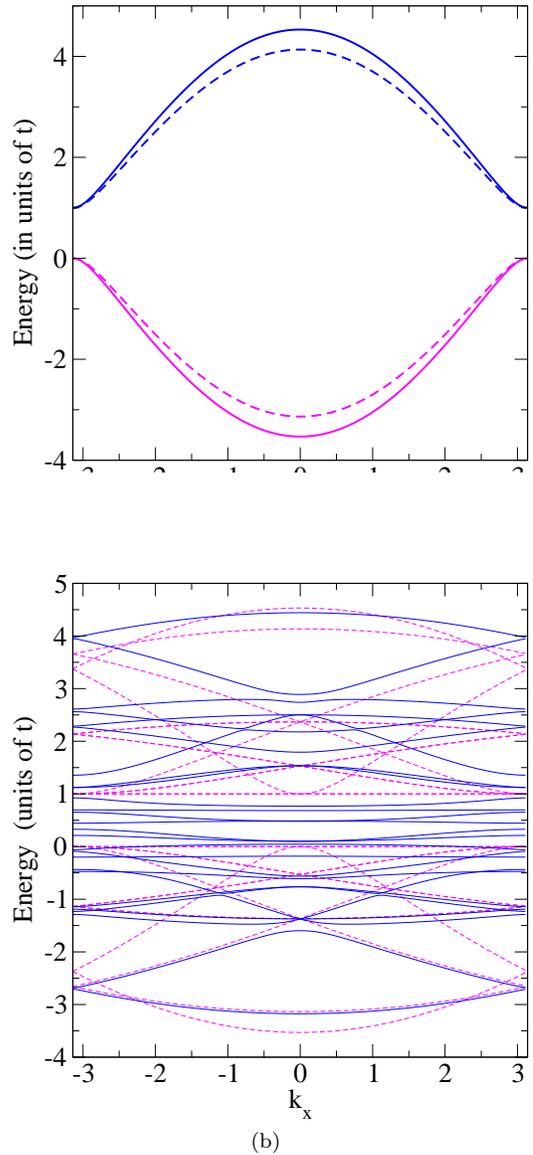

  \begin{center}
%            \subfigure[]{
%              \includegraphics[scale=0.4]{Fig3a.eps} 
%              }\vskip 0.2 in
            \subfigure[]{
              \includegraphics[scale=0.4] {Fig4a.eps}
            }
            \subfigure[]{
              \includegraphics[scale=0.4] {Fig4b.eps}
            }
    \caption{  (a) Bands of the two layers before they
      are placed on top of each other. (b)The band-structure of the two mismatched layers along the $k_x$ direction.}
    \label{fig3}
  \end{center}
\end{figure}

As a 2D example of lattice-mismatched layers,
we consider
a square lattice of atoms $a$ and a different atom $b$ at the center of
every square. (see Fig.~\ref{fig}(a) left). A second layer
of the same structure (Fig.~\ref{fig}(a) right) with the same on-site energies as the first layer, i.e., 
$e^{(a)}_1=e^{(a)}_2= 1$ and $e^{(b)}_1=e^{(b)}_2= 0$ and just a single
hopping $t_{1,2}$ between the atoms $a$ and $b$ of each layer 1 or 2.
Then, the two layers are superimposed to form the periodic super-structure shown
in Fig.~\ref{fig}(b).
As in the one dimensional case in general we can consider the case where $(M+1) a^{(2)} = M a^{(1)}$
where $M$ can be a large integer.
In the example of Fig.~\ref{fig}, a square of 4 unit cells ($2\times 2$) of the layer-1 fits exactly a $3\times 3$ square of layer-2. In this case the corresponding lattice mismatch is very large, i.e.,
the smaller unit cell $a^{(2)}$ is $3/4$ of the larger unit cell $a^{(1)}$.
Without loss of generality, we have considered a rather large mismatch,
which would give
a rather small supercell in order to make its tight-binding treatment
discussed here when
we introduce interlayer hoppings, easier.
With the hopping parameter choice discussed in Appendix~\ref{2d-mismatch} the band-structure
along the $k_x$ direction of the emerging BZ, with an area 4 (9) times smaller than that
of the first (second) layer, is shown in Fig.~\ref{fig3}(b).
The dashed-magenta (solid-blue)
lines correspond to the non-interacting (interacting) layers.
In this example, the interaction between layers is introduced
by using finite hopping matrix elements between layers.
Details on which particular hoppings are introduced and their
values are given in Appendix~\ref{2d-mismatch}.
Notice that allowing for interlayer couplings through interlayer hoppings
some of the degeneracies at the BZ boundaries are lifted opening
gaps (Fig.~\ref{fig3}(b)).

      Notice that the band-structure shares a lot of the characteristics
of our 1D example, with many mini-bands and mini-gaps produced by the
BZ folding and the layer-layer interactions. The interlayer hoppings
used here were rather large which leads to bands inside the original gap.
Smaller values of these parameters will not affect as much the original gap.

\subsection{Twisted bi-layers}
\label{twisted-bilayer}

\begin{figure}[htb]
  \begin{center}
            \subfigure[]{
              \includegraphics[scale=0.4]{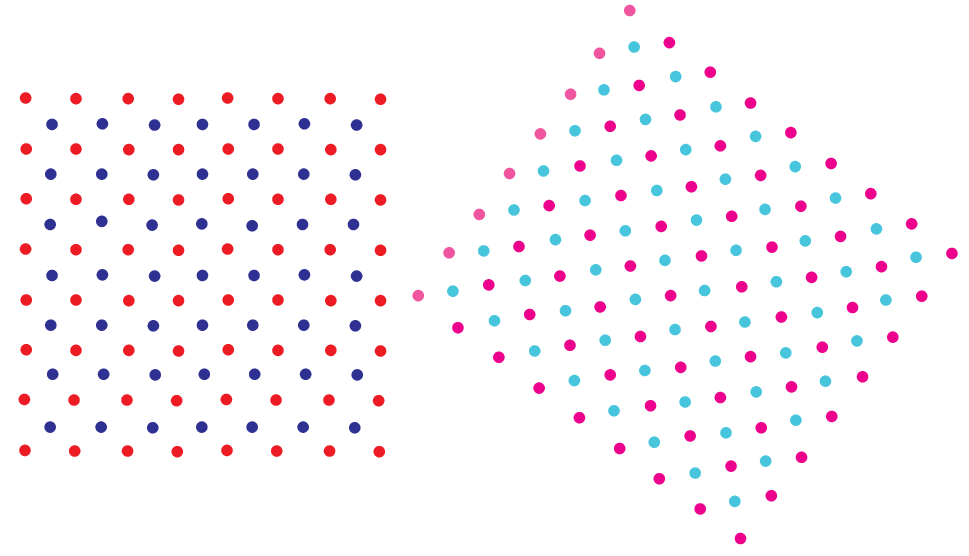}
              }
            \subfigure[]{
              \includegraphics[scale=0.3] {Fig5b.eps}
              }
            \caption{(a) One layer (top left) and a second layer (top right) twisted by an angle
               $\theta$ (in this example $\sin\theta = 4/5$).
              (b) Moir\'e structure produced by placing the two layers
      on top of each other.}
  \label{figx}
  \end{center}
\end{figure}

\begin{figure}
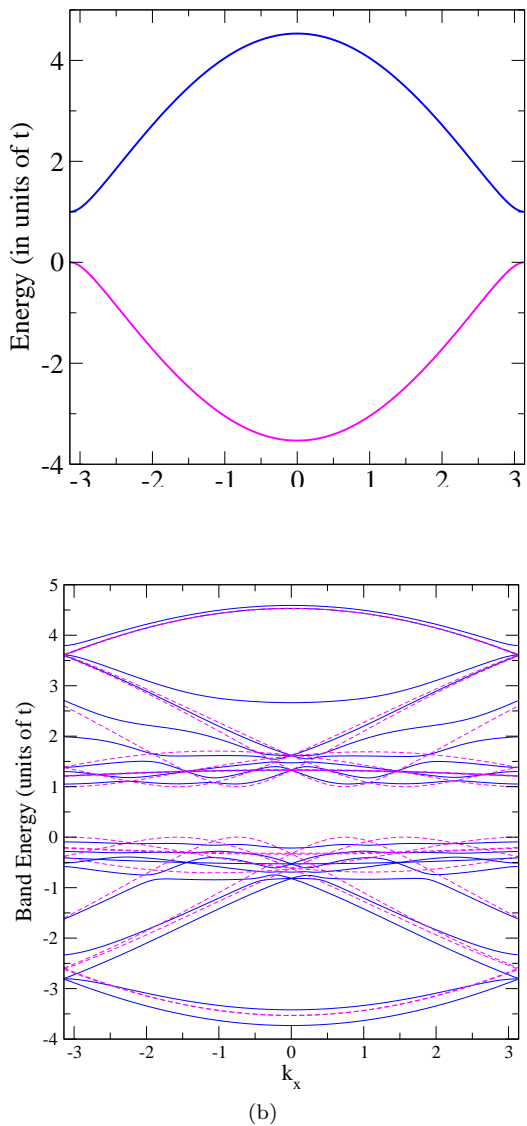

  \begin{center}
            \subfigure[]{
              \includegraphics[scale=0.4] {Fig6a.eps}
            }
            \subfigure[]{
              \includegraphics[scale=0.4] {Fig6b.eps}
              }
    \caption{(a) The band-structure of each of the two untwisted layers.
      (b) The band-structure of the twisted bilayer
      (Magenta (blue) lines denote non-interacting (interacting) layers).}
    \label{fig4}
  \end{center}
\end{figure}
As an example of twisted layers, we considered two atomically-thin layers of the same lattice
discussed in the previous paragraph, i.e., a square lattice of atoms $a$ with
an atom $b$ at the center of the square as shown in the left panel of
Fig.~\ref{figx}(a), and
we twisted one of them by an angle $\theta$ (in this example $\sin\theta = 4/5$)
with respect to the first to obtain the structure shown in the right-panel of
Fig.~\ref{figx}(a). When the rotated layer is placed on top of the
first one, the pattern illustrated in Fig.~\ref{figx}(b) is
produced with a super-lattice constant $a_s=\sqrt{5}a$.

The band-structure along the $k_x$ direction of the small BZ of the
moir\'e super-lattice obtained, as outlined in the Appendix~\ref{twisted-bz}, is illustrated in Fig.~\ref{fig4}(b).
When we include weak interactions (see Sec.~\ref{interactions}), we obtain the blue-lines of
the band-structure along the $k_x$ direction of the small BZ of the
moir\'e super-lattice shown in Fig.~\ref{fig4}(b).
The Brillouin zone of the twisted bilayer is shown in Appendix~\ref{twisted-bz}.
The band structures of each of the untwisted layers is also shown in Fig.~\ref{fig4}(a) for comparison.
The main features (many mini-bands and mini-gaps) of the band-structure
are similar to those of the previous example.

\section{Role of interactions}
\label{interactions}
In this section we study the role of interactions a) on the
tight-binding models introduced in the previous section and b) on the relaxation
of the photo-excited electron-hole pairs.

\subsection{On the tight-binding bands}
\label{interactions:bands}
In the previous Section, we have demonstrated that, as expected, the super-periodic structure formation leads to the folding of the BZ where 
along the mini-BZ edge gaps are expected to form
because the interactions generally lift the degeneracies.

\begin{itemize}
  
\item The simplest form of interactions that can lift these degeneracies
at the BZ edges is when we explicitly introduce non-zero
matrix elements which couple the Bloch state $|n {\bf k} \rangle$ corresponding
to band labeled $n$ with the
Bloch states $|n {\bf k + G} \rangle$ (which in the folded BZ
 will require a different band index) for ${\bf k}$ near the mini-BZ boundary
defined by the reciprocal lattice vectors ${\bf G}$. This approach
was implemented in the mismatched chains in Sec.~\ref{mismatched-chains}
to find the transition from the TB bands to the interacting ones with
gaps formed at the mini-BZ boundaries.
\item Another simple way to include effects of interactions that can create gaps at the mini-BZ boundaries is
when we include interlayer hoppings. This was
demonstrated in our tight-binding model for the lattice mismatched bilayer in
Sec~\ref{mismatched-layers}.
\item The mini-bands have small bandwidth and, therefore, the effects of
  Coulomb interactions are expected to be significant.
  In order to schematically illustrate the role of the Coulomb interactions
on the band-structure as described  within the
quantum lattice gas models defined by our tight-binding models of
the previous section, we will confine our approach to
a mean-field-like approximation.
As an example, we will apply a Hartree-Fock-like approximation to
include some effects of electron-electron interactions in the band
structure of the twisted-bilayer system introduced
in Sec.~\ref{twisted-bilayer}.
\end{itemize}

However, application of the Hartree-Fock approximation\cite{PhysRevB.100.205113} requires knowledge
of the actual atomic orbitals (or Wannier states) in order
to calculate the direct and exchange integrals of the Coulomb interaction.
A generic tight-binding model defined on a given lattice,
only specifies the hopping matrix elements
between such states assumed to live on sites, but not explicitly these states (orbitals) themselves.
To find those one needs to consider
a specific material and carry out a DFT or, in general, a microscopic calculation
with the same goal. Therefore, we will have to extend our TB model to one where
the interactions are introduced also at the level of another model (such an example is
the Hubbard model).

Our tight-binding model is a quantum lattice-gas model which consists of
sites inside a supercell with index ${\bf R}_s$. Inside each such supercell
$N_o$ orbitals, specified with the index $\alpha=1,...,N_0$, reside. After the TB
calculation the Block states are written as
\begin{eqnarray}
  | \Psi_{n,\bf k} \rangle = \sum_{\alpha=1}^{N_0} c^{(n)}_{\alpha}({\bf k})
  | {\bf k}, \alpha \rangle,
\end{eqnarray}
where
\begin{eqnarray}
  | {\bf k}, \alpha \rangle =
  {1 \over {\sqrt{N_s}}} \sum_{{\bf R}_s}
  e^{-i {\bf k}\cdot {\bf R}_s} | {\bf R}_s, \alpha\rangle,
  \label{wannier}
\end{eqnarray}
and $| {\bf R}_s, \alpha\rangle$ represents the $\alpha$ orbital
in the supercell specified by the lattice vector ${\bf R}_s$.
The coefficients $c^{(n)}_{\alpha}({\bf k})$ are found after the diagonalization
of the $N_0\times N_0$ TB matrix for a given value of ${\bf k}$ of
the mini-BZ.

The Hartree-Fock Hamiltonian written in the basis of the atomic or (Wannier)
states
given by Eq.~\ref{wannier} in momentum space has the following form:
\begin{eqnarray}
  H_{\alpha\beta} ({\bf k}) &=& T_{\alpha\beta}({\bf k}) + V_{\alpha\beta}({\bf k}),
    \\
  V_{\alpha\beta}({\bf k}) &=&    \sum_{\gamma,{\bf k}'}
  n_{\gamma}({\bf k}')
  \Bigl (U_{\alpha {\bf k} \gamma {\bf k}'\gamma {\bf k}'\beta {\bf k}} - U_{\alpha {\bf k} \gamma {\bf k}' \beta {\bf k} \gamma {\bf k}'} \Bigr ),      \label{hartree-fock}
  \end{eqnarray}
where
\begin{eqnarray}
  U_{\alpha_1 {\bf k}_1 \alpha_2 {\bf k}_2\alpha_3 {\bf k}_3\alpha_4 {\bf k}_4}&=&
  \langle {\bf k}_1, \alpha_1, {\bf k}_2 \alpha_2 | U_c|
   {\bf k}_3, \alpha_3, {\bf k}_4 \alpha_4 \rangle
\end{eqnarray}
is the two-body matrix element of the Coulomb interaction between
pairs of Wannier states given by Eq.~\ref{wannier} and Eq.~\ref{hartree-fock}
includes both the direct and the exchange terms.
Here $T_{\alpha\beta}({\bf k})$ is the TB matrix which was diagonalized
in the previous section in order to obtained the non-interacting folded bands.
The summation over $\gamma$  is over occupied states, i.e., $n_{\gamma}({\bf k})$ is the Fermi-Dirac distribution.
The starting tight-binding model is a general model in which only the
hopping matrix elements between sites are specified and the atomic
or more generally the Wannier orbitals
that correspond to the sites are not specified. 
Therefore, instead of starting by
making a model for these Wannier states and, then, calculating the
matrix elements of $U_c$, we are going to make a model
for the interaction matrix elements $V_{\alpha\beta}({\bf k})$
directly.
Thus, in order to schematically model the role of interactions on the band structure, we have used
the following two different forms for such matrix elements:
\begin{eqnarray}
V_{\alpha\beta}({\bf k}^{\prime}) &=& V_0 \exp \Bigl [-\Bigl ({{E^{(0)}_{\beta \bf k} - E^{(0)}_{\alpha \bf k}} \over {\Gamma}} \Bigr )^2 \Bigr ],\\
V_{\alpha\beta}({\bf k}) &=& {{V_0} \over {1+\Bigl ({{E^{(0)}_{\beta \bf k} - E^{(0)}_{\alpha \bf k}} \over {\Gamma}} \Bigr )^2}}.
\end{eqnarray}
Here $E^{(0)}_{\alpha \bf k}$ are the TB (non-interacting) energy bands and
$V_0$ and $\Gamma$ are energy scales. Both models state the expectation that
electrons in bands which are close in energy interact more strongly
and the parameter $\Gamma$ gives the range of this interaction.
We find that for small values of $V_0$ 
and/or large values of $\Gamma$ both models give essentially the same results.
Depending on how strongly or weakly localized the Wannier orbitals are,
the dependence of these
matrix elements on the band-energy difference can be  weak or strong.
Another factor which will influence this dependence is the screening
effects as reflected in the dielectric function of the material.

\iffalse
Before we apply this approach
to the twisted-bilayer problem, we would like to note that this Hartree-Fock
approach is similar to the method where we introduce matrix elements to
couple the otherwise non-interacting bands.
\fi

\begin{figure*}
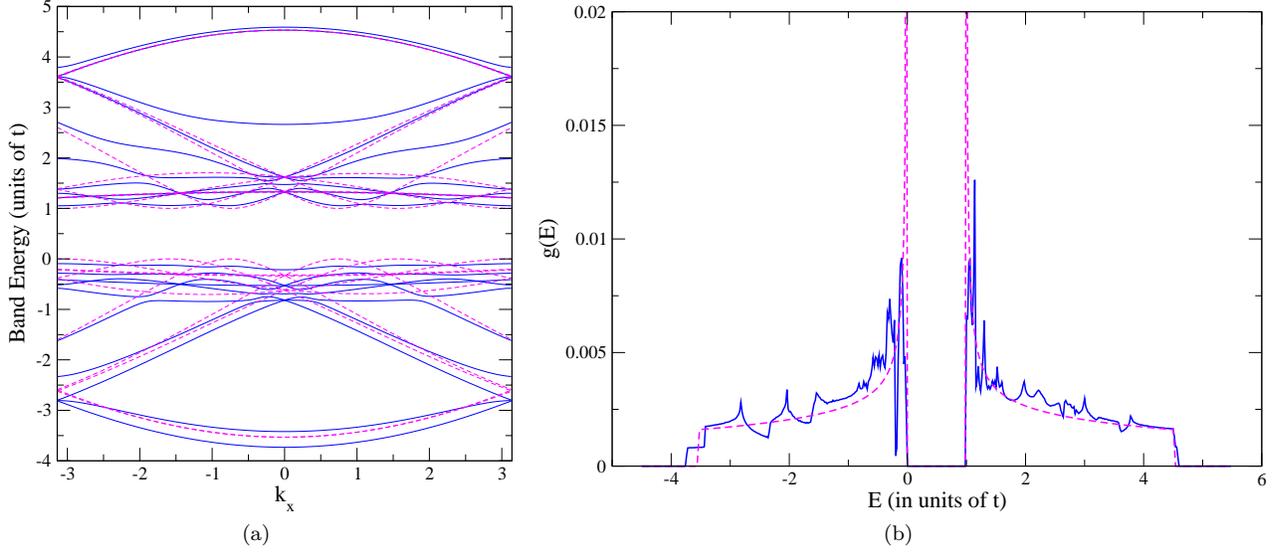

  \begin{center}
            \subfigure[]{
              \includegraphics[scale=0.4]{Fig7a.eps} 
              \label{bands-int}
              }
            \subfigure[]{
              \includegraphics[scale=0.4]{Fig7b.eps}
              \label{ds-int}
            }
            \caption{(a) The bands of the twisted bilayer after the
              inclusion of the interaction discussed in the text for
              $V_0/t=0.1$. (b) The density of states of the interacting
              twisted bilayer (blue solid line) is compared to that of
              the non-interacting (magenta dashed-line).
}
  \end{center}
\end{figure*}

\begin{figure*}
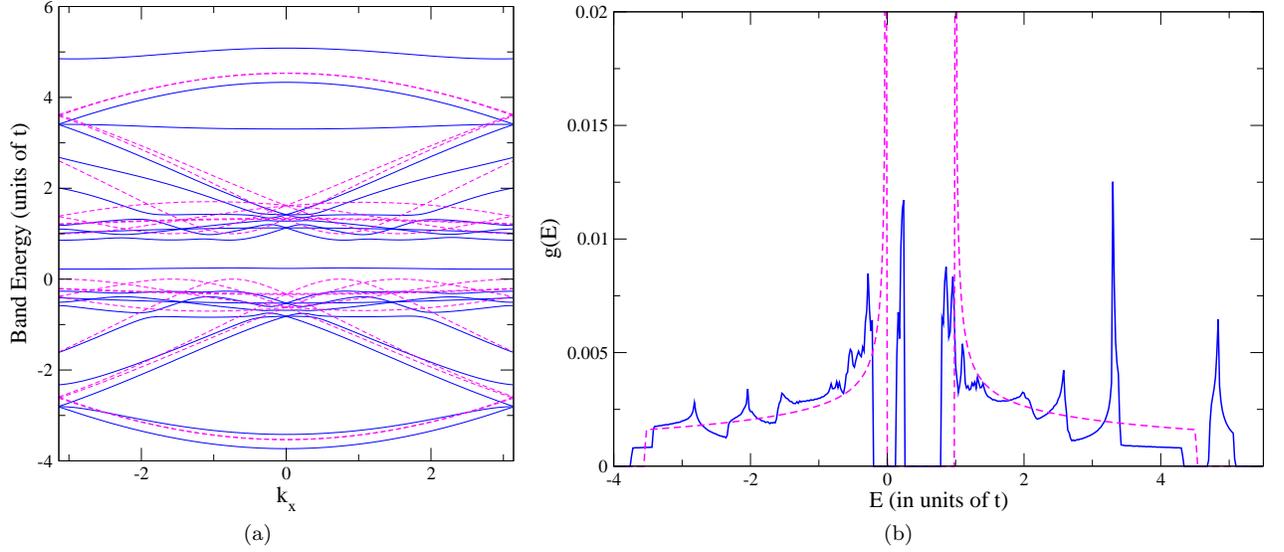

  \begin{center}
            \subfigure[]{
              \includegraphics[scale=0.4]{Fig8a.eps} 
              \label{bands-int0.2}
              }
            \subfigure[]{
              \includegraphics[scale=0.4]{Fig8b.eps}
              \label{ds-int0.2}
            }
            \caption{(a) The bands of the twisted bilayer after the
              inclusion of the interaction discussed in the text for
              $V_0/t=0.2$. (b) The density of states for $V_0/t=0.2$ (blue solid line) is compared to that of
              the non-interacting (magenta dashed-line).
}
  \end{center}
\end{figure*}

We have carried out the diagonalization of our HF-like model. Notice that in the bands presented in Fig.~\ref{bands-int} the interaction opens gaps near the mini-BZ edges. However, not all the degeneracies at
the BZ boundaries are lifted within the HF approximation and some
will remain in some ``theoretically exact'' treatment.
Some of these degeneracies should be
lifted through the electron-phonon interaction
by allowing the superlattice to relax.
If the interactions are weak, the gap-openings take place only in the
vicinity of the mini-BZ edges and do  not become global gaps,
i.e., through
the entire BZ; they only lead to a depression of the density  of states (DOS)
in the vicinity of the band energy at the mini-BZ edges as illustrated in
Fig.~\ref{ds-int}. 
Some gaps form in the DOS, but more generally  a depression in DOS appears.
Our DOS share very similar features
as the HF calculation of  Ref.~\cite{PhysRevB.100.205113} for twisted
bilayer graphene, with the notable difference that graphene has
no gap while our model for a general semiconductor parent material
does have a gap.
When the interaction is stronger as shown in Fig.~\ref{bands-int0.2}
for $V_0/t=0.2$ the gaps at the mini-BZ edges become larger and as shown in Fig.~\ref{ds-int0.2}  the depressions of the DOS
near the band energy of the mini-BZ edges becomes deeper.

In the following subsection, we will argue that these small (partial) gaps
lead to enhancement of the PV efficiency because they limit the phase
space for relaxation of the photo-excited electron-hole pairs via phonon
emission. This reduces the rate of phonon emission which allows 
the impact ionization processes to take place as discussed in the next
subsection. 

  \subsection{On the relaxation of the photo-excitations}
\label{interactions:relaxation}

Now, let us consider the various interaction processes which
take place in the photo-excited electron/hole system in
an insulator with a band-structure in which the conduction
and/or the valence bands have been fragmented by a multitude 
of small partial or full energy gaps created by either of the mechanisms discussed in the previous subsection.

\iffalse
To be clear, the above two processes can also take place in a material
with large bandwidth conduction and/or valence band as in Fig.~\ref{fig3}(a).
In such case, however, the process of phonon emission can continue until the electron ends up in the
bottom of the band minimum, if the impact-ionization in the
visible spectrum is smaller than the rate of phonon-emission. \fi

We will consider the role of the interaction between the
photo-excited electrons and
holes with a) phonons, b) other electrons and holes,  and
c) with the incident solar photons.
These interactions lead to a dynamic quasi-equilibrium state which
leads to a finite photo-current. We are ultimately interested in
finding the efficiency of converting the energy of the solar
radiation into electrical energy.

\begin{figure}
  \begin{center}
            \subfigure[]{
              \includegraphics[scale=0.25]{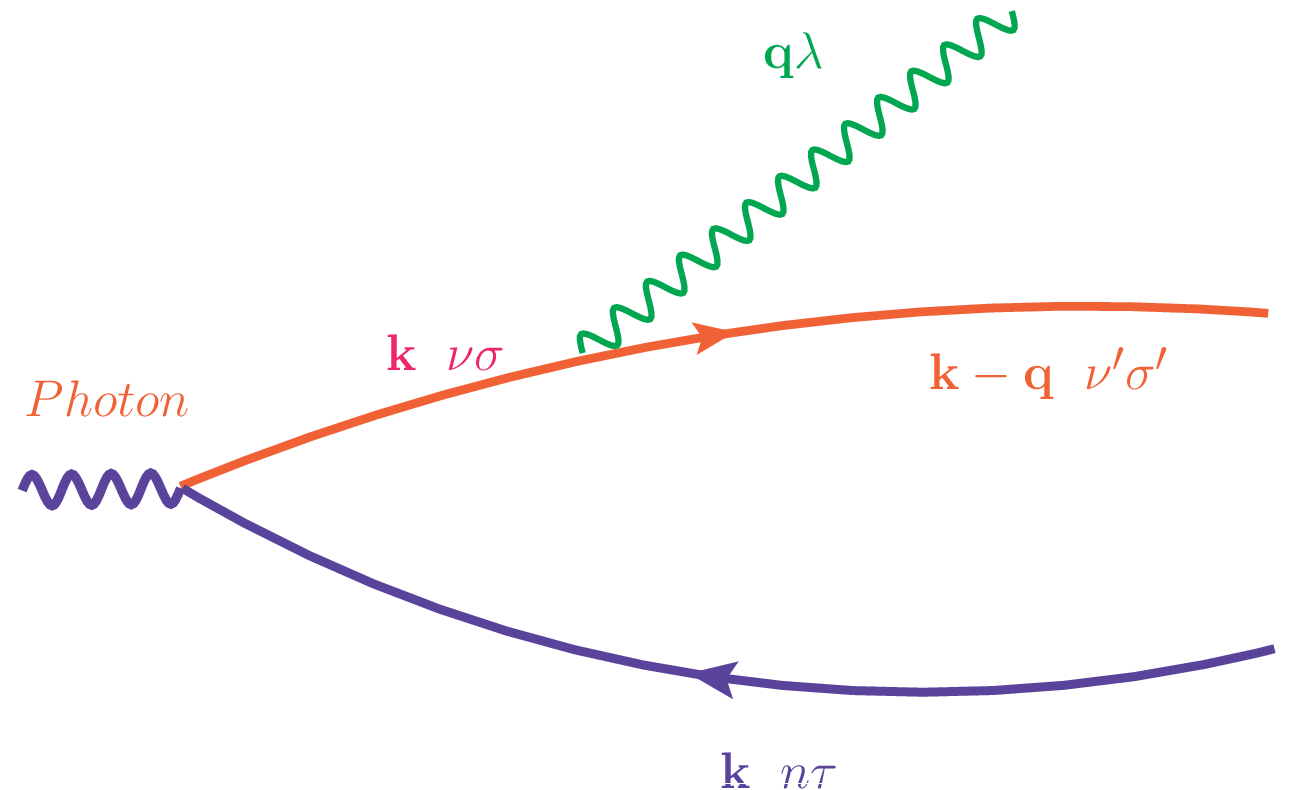} 
              \label{phonon-emission}
            } \\
            \vskip 0.5 in
            \subfigure[]{
              \includegraphics[scale=0.25]{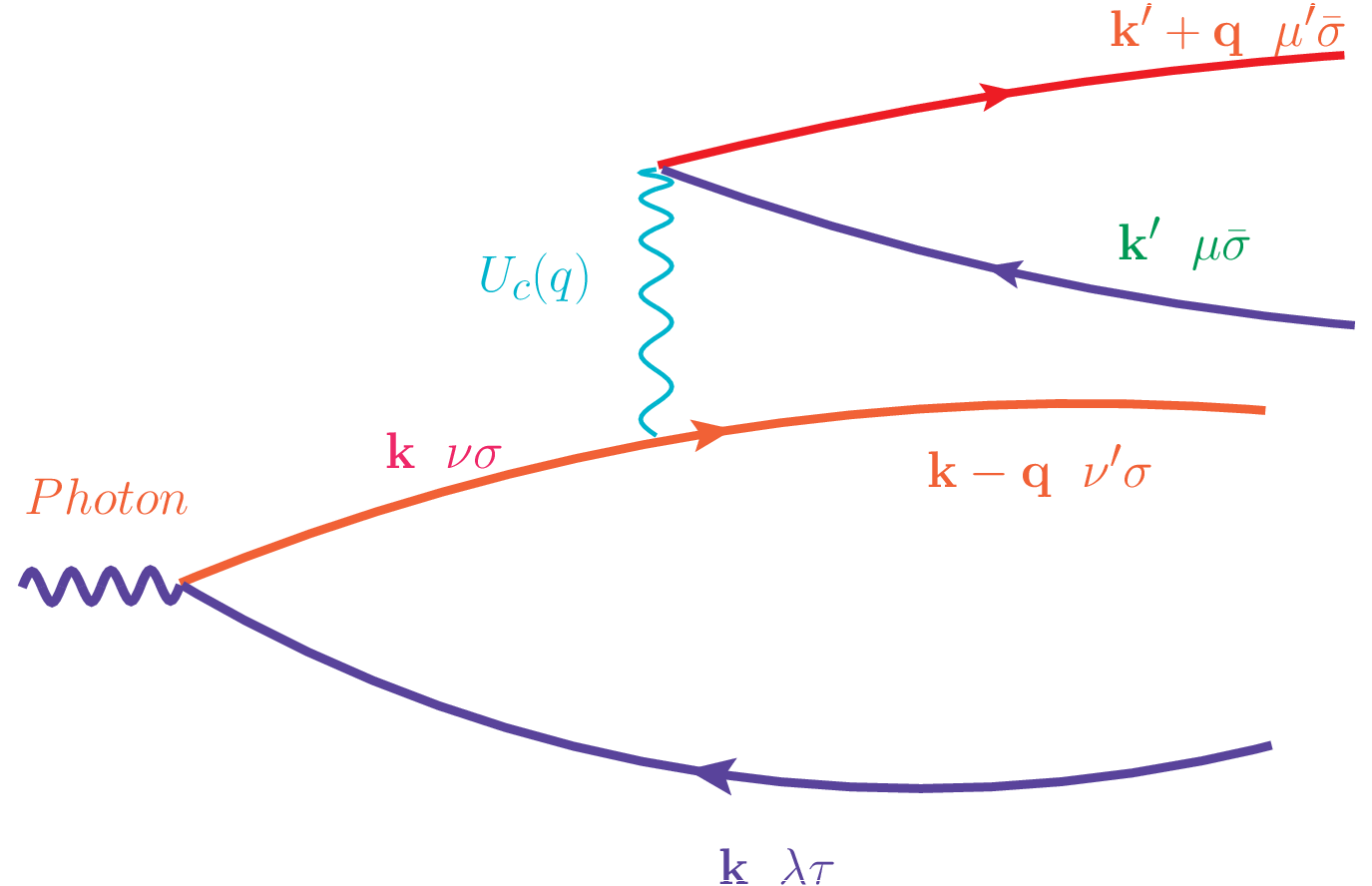} \label{impact}
              }
            \caption{(a) Within the photo-excited electron-hole pair the electron emits
              a phonon and, thus, decays from the state $\nu {\bf k}\sigma$
              to the state $\nu' ({\bf k-q})\sigma'$ by emitting a phonon
              of momentum ${\bf q}$, polarization $\lambda$ and frequency $\omega^{(\lambda)}_{\bf q}$
              (here $\nu$ and $\nu '$ are band labels).
              (b) Within the photo-excited electron-hole pair the electron 
interacts with a different valence electron and it promotes it to the conduction band, thus, leaving a hole behind.  This is the process of the so-called impact ionization.}
  \end{center}
\end{figure}

First, the photo-excited electron/hole pairs can transfer energy to phonons
through the phonon-emission process illustrated in Fig.~\ref{phonon-emission}
or a cascade of such processes:
\begin{eqnarray}
{\rm {e}} \leftrightarrow {\rm {e}}^{\prime} + {\Gamma},
\label{decay-to-phonon}
\end{eqnarray}
where $\Gamma$ denotes the emitted phonon of momentum $\vec q$ and polarization $\lambda$. The initially photo-excited electron state $e$
decays to another electronic state $e^{\prime}$ of a different energy and
momentum by emitting a phonon or multi-phonons.
The decay rate of such processes can be estimated by
\begin{eqnarray}
  { 1 \over {\tau^{phonon}_{\vec{k} n}}} &=&
  \sum_{\vec q,\lambda} |M_{\nu\nu'\lambda}(\vec k,\vec q)|^2 \nonumber \\
  &&\delta(\omega - (E_{\nu}(\vec k)-E_{\nu'}(\vec k-\vec q) - \hbar
      \omega^{(\lambda)}_q),
\end{eqnarray}
where $M_{\nu\nu'\lambda}(\vec k,\vec q)$ is the electron-phonon interaction vertex and $E_{\nu}({\bf k})$, and $\omega^{\lambda}({\bf q})$ are the electron-band energy and
phonon frequencies.
This decay rate of the excited electron or hole when $\omega$ is within the spectrum of the solar radiation is of the order of $10^{12}-10^{13}$/sec for the typical case of Si as calculated
from first principles\cite{Si-phonon-Louie}. As we shall see this can be much smaller than the impact ionization rate (IIR) as the latter involves
much larger energy scales related to the much stronger Coulomb interaction.
Even when the created small gaps $\delta G$ are of the order of 0.05 eV or even
somewhat larger
in the insulators considered here, they can be sufficiently large to reduce or
stop the decay via the process described by Eq.~\ref{decay-to-phonon}
and illustrated in Fig.~\ref{phonon-emission}. The reason is that the
phase space which allows such transitions gets significantly reduced because
the phonon energies $\hbar \omega^{(\lambda)}_q$ are smaller (less than $\sim 0.05$
 eV) to that needed to take enough energy from the electron
 to allow it to decay to a lower electronic state separated from its present
 state by this small gap, because 
$\hbar \omega^{(\lambda)}_q < \delta G$. The vertex (matrix element) of an electron emitting multiple
phonons at the same time is very small, therefore, the decay has to be done
with successive single phonon emission processes and each one will have a probability significantly less than unity because of the small gap.

Another process of decay is the so-called impact ionization
process (IIP) shown in Fig.~\ref{impact}, which increases the number of charge carriers per absorbed
phonon.
In order to calculate the IIP rate we need to include the processes
illustrated
diagrammatically in  Fig.~\ref{impact}, where
the electron or the hole component of the photo-excited
electron-hole pair decays into  two-electron-one-hole or two-hole-one-electron
states respectively\cite{PhysRevB.82.125109,Coulter2014}.
An estimate of the IIP rate can be obtained  using the
quasiparticle self-energy obtained from quasiparticle-self-consistent GW (QscGW) calculations.
Namely, the IIP rate $\tau_{\vec{k} n}^{-1}$ is calculated from the self-energy
$\Sigma_{\vec{k} n}(\omega)$ as follows:
\begin{eqnarray}
{ 1 \over {\tau_{\vec{k} n}}} &=& \frac{2 Z_{\vec{k} n}}{\hbar} 
                        | \rm{Im} \, \Sigma_{\vec{k} n}(\omega) |, \\
Z_{\vec{k} n} &=& \left(1 - \rm{Re} 
                \frac{\partial \Sigma_{\vec{k} n} (\omega)} 
                {\partial \omega}\Big|_{\epsilon_{\vec{k} n}} \right )^{-1},
\end{eqnarray}
where $\vec{k}$ and $n$ are k-point and band indices. When $n$ band 
indices are running for valence bands, then, they correspond to the 
recombination to the hole-initiated biexcitons, otherwise to the electron-initiated biexcitons.
For the self-energy calculations, we use the same parameters as in the 
QscGW calculations mentioned earlier.  This rate has been calculated\cite{Coulter2014} for insulators with flat
bands, where the magnitude of the Coulomb interaction is much larger than their
bandwidths, and it was found to be of the order of $10^{15}$/sec in the spectrum of the
solar radiation which is larger by 2 orders of magnitude as compared to the
decay rate through phonon emission\cite{Si-phonon-Louie}. Therefore, IIP can dominate the energy relaxation processes in such materials.
More importantly, the bands discussed in the previous section, have bands with bandwidths small compared to 
the magnitude of the Coulomb interaction because of the band-folding. Therefore, one expects an significantly increased IIP rate\cite{Coulter2014}.

However, Auger recombination which is the
inverse process should not be omitted.
To analyze that, first, let us begin  from the open-circuit situation.
We have found\cite{Manousakis2019} that for the type of insulators discussed
in the previous section with flat bands, the characteristic time
scale for thermalization of the photo-excited electrons
and holes is much shorter than the time needed for 
phonons to thermalize the electronic system with the lattice.
Therefore, it is reasonable to assume 
that the photo-created electrons and the holes in these insulators
thermalize among themselves at a temperature $T_e$ higher than
the lattice temperature $T_L$, by means of the electron-electron
interaction.  This interaction between the
electronic degrees of freedom leads to processes such as
impact ionization and Auger recombination, as well as
carrier-carrier scattering. This equilibrium state is achieved
as follows.

The quasi-equilibrium between the impact ionization process shown in Fig.~\ref{impact} and  the Auger recombination process, i.e., the inverse
of the impact ionization process, shown as the reverse process
of the one shown in Fig.~\ref{impact}, can be denoted as follows 
\begin{eqnarray}
{\rm {e}}_1 \leftrightarrow {\rm {e}}^{\prime}_1 + {\rm {e}}_2 + {\rm h},
\label{impact1}
\end{eqnarray}
where e$_1$ is the initial electronic state which interacts
with a valence electron and promotes it to the
conduction band, thus, creating an additional e-h pair denoted as
e$_2$-h, by changing its energy and momentum to become the state e$^{\prime}_1$.
In addition, there is the analogous processes for high energy holes, i.e.,
\begin{eqnarray}
{\rm {h}}_1 \leftrightarrow {\rm {h}}^{\prime}_1 + {\rm {e}} + {\rm h}_2.
\label{impact2}
\end{eqnarray}
Last, the
electron-electron scattering due to the Coulomb repulsion 
 leads to electron-electron, hole-hole, and electron-hole interaction processes, which lead to the following quasi-equilibrium equations:
\begin{eqnarray}
{\rm {e}}_1 + {\rm {e}}_2 &\leftrightarrow&  {\rm {e}}^{\prime}_1
+ {\rm {e}}^{\prime}_2, \label{electron-electron1}
\\
{\rm {h}}_1 + {\rm {h}}_2 &\leftrightarrow&  {\rm {h}}^{\prime}_1
+ {\rm {h}}^{\prime}_2, \label{electron-electron2}
\\
{\rm {e}} + {\rm {h}} &\leftrightarrow&  {\rm {e}}^{\prime}
+ {\rm {h}}^{\prime}.
\label{electron-electron3}
\end{eqnarray}
If these scattering processes are excluded, and we consider only
impact ionization and Auger recombination, 
then, only those processes involving electrons and holes in certain energy
and momentum range (which allow for impact ionization and its
inverse through energy-momentum conservation) would take place
and everything else would be excluded. As a result, the final 
distribution of carriers would depend on energy and momentum and, thus,
it cannot
be described by a Fermi-Dirac distribution. We would like to remind the
reader that we have neglected the
effects of 
electron-phonon interaction because, as we have shown, they lead to equilibrium with
the lattice at a temperature $T_L$ at  a much longer time-scale. The above
electronic scattering processes are important in order to establish a quasi-equilibrium of the electronic degrees of freedom
at a common temperature $T_e$ much higher than $T_L$
and the electron/hole distribution becomes a Fermi-Dirac distribution with
characteristic temperature $T_e$. Therefore, these scattering processes
are absolutely important and we include them as part of the mechanism
which leads to the fast electronic quasi-equilibrium.

\iffalse
In the next section, the equilibrium of these processes together
with the photo-excitation process:
\begin{eqnarray}
\gamma \leftrightarrow {\rm {e}} + {\rm {h}}
\label{photo-excitaton}
\end{eqnarray}
are considered together in order to find the 
equilibrium between the absorbed and emitted energy currents, which leads
to the PV efficiency.
\fi
\section{Equilibrium and PV-efficiency}
\label{sec:efficiency}

We consider a ``parent'' bulk material  that consists of valence and conduction bands with
bandwidths $W_v$ and $W_c$ separated by a relatively large gap
$E_g$. In addition, $W_v$ and $W_c$ are larger than 3.5 eV, the approximate
upper bound of the solar radiation spectrum. We will assume that, through
either of the two mechanisms discussed in the previous section (lattice mismatched or twisted layers),
these large bands split into many bands with smaller band-widths
$\delta W$ and small gaps $\delta G$.
In order to be concrete we consider a simplified but quite general model for the energy bands.  We will assume that the overall density of states
of the parent bulk semiconducting material used to make these
superlattice structures does not change, only small gaps open, creating the following
energy bands and gaps:
\begin{eqnarray}
  E^{(b)}_n &=& E_g + (n-1)(\Delta W + \delta G),\\
  E^{(t)}_n &=& E^{(b)}_n + \Delta W.
  \label{simplified-bands}
\end{eqnarray}
where $n=1,2,...,M$. First, there is a relative large gap $E_g$, i.e., the gap in the original
bulk semiconductor. The band-folding, due to  the
super-structure formation from the lattice mismatch or the moir\'e
periodicity, causes the opening of small gaps $\delta G$ which split the original conduction band. 

This model is a simplified version of what actually happens in the
actual material as discussed in the previous section. However, here,
what we wish
to show is the following: These additional gaps prevent
absorption of the solar radiation when the energy of the incident photon falls
within these gaps. Therefore, naively, one would expect a reduced PV efficiency.
However, as we argued in the previous section these small gaps  prevent energy loss via phonon emission and enhance the  rate of IIP.
Next, we will show that the overall effect is
to significantly enhance the PV efficiency, especially in the limit
of very small gaps.

To begin, the  flux of the incident solar radiation
creates the energy current of the absorbed solar photons
given by
\begin{eqnarray}
J_{absorbed} = {{\Omega_s} \over { 4 \pi^3 \hbar^3 c^2} }
\sum_{n=1}^M \int_{E^{(b)}_n}^{E^{(t)}_n}
d\epsilon {{\epsilon^3} \over {\exp({{\epsilon} \over {k_B T_s}}) -1}},
\end{eqnarray}
where  $\Omega_s$ is the solid angle under which the Sun is seen, 
and $T_s$ is the
Sun's surface temperature $T_s\simeq 5760$ K. Here, the incident radiation is
not absorbed when the photon energy falls within any of the band gaps.

If we now consider the equilibrium of all of the 
processes discussed in the previous section, we may realize that  they produce 
a distribution of electron/hole pairs in quasi-equilibrium with the distribution of
photons over energy due to the emission process:
\begin{eqnarray}
e + h \leftrightarrow \gamma.
\label{emission}
\end{eqnarray}
in which an electron (e) and a hole (h) combine and a photon ($\gamma$) is emitted 
(luminescence) or the inverse where a photon produces an electron/hole pair
in these  insulators.
This distribution of photons in equilibrium with 
a gas of quasiparticles consisted of electrons and holes 
can be described  by a thermal
distribution of photons at a temperature $T_e$ and
a chemical potential $\mu_{\gamma}=\mu_{e-h}=0$. The value
of $T_e$ is determined from the equation
\begin{eqnarray}
J_{emitted} &=& J_{absorbed},\label{equilibrium}
 \\
J_{emitted} &=& {{\Omega_e} \over { 4 \pi^3 \hbar^3 c^2} }
\sum_{n=1}^M \int_{E^{(b)}_n}^{E^{(t)}_n}
d\epsilon {{\epsilon^3} \over {\exp({{\epsilon-\mu_{\gamma}} \over 
{k_B T_e}}) -1}}.
\end{eqnarray}
Here $\Omega_e$ is the solid angle of the emitter 
and it is $\pi$ for a planar emitter.
This equation defines a temperature $T_e$ of the electronic
system which is considered decoupled from the lattice and the phonons.
A value of $\mu_{e-h}=0$ is expected for a system in which the particle
number is not conserved as is the case for the electron and holes in the 
equilibrium state determined by the impact ionization and Auger-recombination.

If we assume that the density of the electrons and of holes near the interface
of the PV junction
is not large, the above process produces an quasi-equilibrium 
Fermi-Dirac distribution of electrons and holes at the above temperature
$T_{e}$ with no separation of the quasi-Fermi-energies, i.e, the chemical
potential difference between electron and holes $\mu_{e-h}$ is zero.

Under open circuit conditions, this quasi-equilibrium is achieved
 at the
temperature $T_e$, which is higher than the lattice temperature
$T_L$. As discussed in the previous section, the time scale 
in these insulators to reach equilibrium (through impact ionization,
 Auger recombination and electron-electron scattering) is much 
shorter that the time scale required for such a system to reach thermal
equilibrium with the lattice through the process described by Eq.~\ref{decay-to-phonon} and the following process
\begin{eqnarray}
e + h \leftrightarrow \Gamma,
\label{phonon}
\end{eqnarray}
where $\Gamma$ denotes a phonon. This happens because
\begin{itemize}
\item the time-scale for the decay processes described by 
Eqs.~\ref{impact1} and ~\ref{impact2} is much faster than the process
of electron decay via phonons and

\item the interaction matrix elements
involved in  the processes described by Eq.~\ref{electron-electron1},
\ref{electron-electron2},\ref{electron-electron3}
are much stronger than the electron-phonon interaction matrix
elements leading to the processes \ref{phonon}.
\end{itemize}

These facts allow first equilibration of the electronic system 
to a temperature $T_e$, which is different than $T_L$ by creating
multiple electron-hole carriers per incident photon. This phenomenon has been argued
in Refs.~\onlinecite{PhysRevLett.72.3851,BRENDEL1996419,PhysRevB.52.11319} to lead to a significant increase of
the efficiency by means of carrier multiplication \cite{BRENDEL1996419}
for a conventional solar cell.

With these facts in mind, we treated the carriers in the insulator as weakly
interacting quasiparticles which obey Fermi-Dirac statistics, which may allow
us to apply the same assumptions and approximations used by Brendel 
et al.\cite{BRENDEL1996419} to calculate the efficiency of a solar
cell made from the material characterized by the band structure described by
Eq.~\ref{simplified-bands}. We find that the maximum efficiency is bounded from above by 
the maximum of the following function of the voltage $V$ and the gaps $\delta G$:
\begin{eqnarray}
  \eta(V) &=&   q V {{ G-   R(V)} \over {P_{in}}},
\end{eqnarray}
where $q$ is the carrier charge and we are seeking the maximum of $\eta(V)$
with respect to the voltage $V.$ $P_{in}$ is the
incident solar radiation power  given by
\begin{eqnarray}
P_{in} &=& \kappa_s  \int_{0}^{\infty} d\epsilon 
{{\epsilon^3} \over {\exp(\epsilon/k_BT_s)-1}},
\end{eqnarray}
where the $\kappa_s = e_s { 2 \over {c^2 \hbar^3}} $
and $e_s = \pi A \sin^2\phi_s $ is the illumination \'etendue\cite{10.1063/1.1736034} of a cell of area $A$ that sees the Sun under a half angle $\phi_s$. 
As already mentioned, $T_s$ is the temperature on the solar surface. 

The quantity $G$  is the generation current
given by
\begin{eqnarray}
G &=& \kappa_s \sum_{n=1}^{\infty}\int_{E^{(b)}_n}^{E^{(t)}_{n}} d\epsilon  
{{m(\epsilon) \epsilon^2} \over {\exp(\epsilon/k_BT_s)-1}},
\end{eqnarray}
$m(\epsilon)=\min([{{\epsilon} \over {E_g}}],m_{\mathrm{max}})$ is the number of
electron-hole pairs\cite{BRENDEL1996419} generated by one hot carrier of energy $\epsilon$ through the process of impact ionization, 
where $[{{\epsilon} \over {E_g}}]$ stands for the integer part
and $m_{\mathrm{max}}$ is a maximum  allowed value of the carrier multiplication.

$R(V)$  is  the electron-hole recombination current which produces  photons or due to
the Auger process current (the inverse process of the process illustrated
in Fig.\ref{impact}). The result is a recombination current given as
\begin{eqnarray} 
R(V) &=& \kappa_s \xi  \sum_{n=1}^{\infty} \int_{E^{(b)}_n}^{E^{(t)}_{n}} d\epsilon 
{{ m(\epsilon) \epsilon^2} \over {\exp((\epsilon-q V)/k_BT_e)-1}},
\end{eqnarray}
where  $\xi = \pi/\Omega_s$  and
$\xi=46396$ ($\xi=1$) corresponds to unconcentrated (fully concentrated) sunlight.

\begin{figure}
  \begin{center}
    \includegraphics[width=3. in]{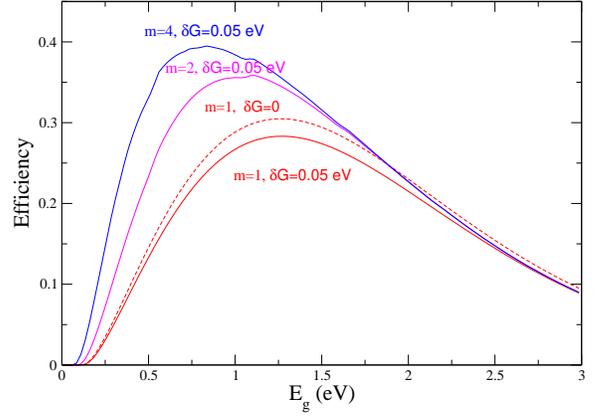} 
    \caption{Calculated efficiency  ($\delta G=0.05 eV$, $\delta E = 0.5 eV$)  for unconcentrated light.  Here $m$ is the number of electron-hole pairs
      allowed to be generated by a single solar photon (when it has sufficient
    energy) through the process of impact ionization (Fig.~\ref{fig5}(b-c)).}
  \label{fig6}
  \end{center}
\end{figure}

In Fig.~\ref{fig6}, the calculated efficiency is shown for
$\delta G=0.05 eV$ and $\delta E = 0.5 eV$ for the case of unconcentrated sunlight for various values of $m$, which is the number of electron-hole pairs
      allowed to be generated by a single solar photon when it has sufficient
      energy through the process of impact ionization
      (see Ref.~\cite{BRENDEL1996419}). The red-dashed line in Fig.~\ref{fig6} shows the
      result of the $\delta G=0$ and $m=1$ which yields the 
Shockley-Queisser      (SQ) result.
Notice that the corresponding result for $m=1$ but with $\delta G=0.05$ eV
is somewhat lower than the one for $\delta G=0$, as expected as
some photons go through the gaps. However, splitting the
bands allows for impact-ionization to occur, i.e., $m$ can be greater than 1.
As a result we see that regardless of the fact that some photons
are going through the gaps, for $m=2$ (magenta) and $m=4$ (blue)  the efficiency is
significantly higher that the SQ result. Therefore, this calculation
contrary to naive expectations strongly suggests that opening small gaps in the electronic spectrum can enhance the PV efficiency. Namely, while the PV cell
would miss those photons which have the energy to go through the gap,
the net efficiency increases because these gaps allow for the
impact ionization process to occur.

\begin{figure}
  \begin{center}
        \includegraphics[width=3. in]{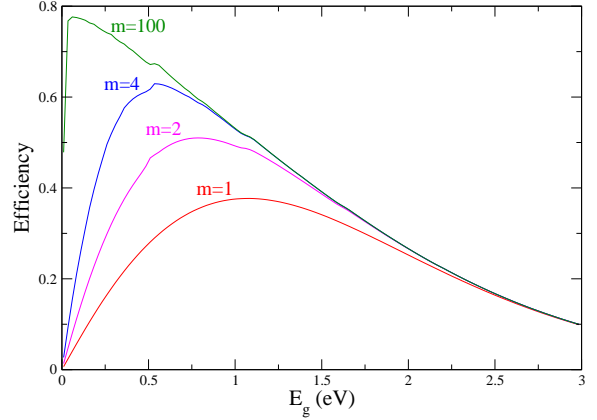}
    \caption{Calculated efficiency  ($\delta G=0.05 eV$, $\delta E = 0.5 eV$)  for fully concentrated sun-light.  }
  \label{fig7}
  \end{center}
\end{figure}

In Fig.~\ref{fig7}, the calculated efficiency is shown for
$\delta G = 0.05 eV$ and $\delta E = 0.5 eV$ for the case of fully concentrated sunlight ($\xi=1$) and for various values of $m$.
Notice the dramatic increase in efficiency due to carrier
multiplication in this case, especially in the limit of vanishingly small
gap and allowing for the impact ionization process to work for arbitrarily
large but energetically possible values of $m$.

\section{Discussion and Conclusions}
\label{sec:conclusions}
We have calculated the PV efficiency enhancement when a semiconductor band
splits into multiple bands with small bandwidths
separated by small gaps by extending the work of
Shockley-Queisser\cite{10.1063/1.1736034} and by including the effects of impact ionization
which as we show becomes much more probable in this case. Realization of this case can be
found in alternating-layer structures 
of lattice mismatched  or misaligned (twisted) atomically-thin layers.
We show that there is a significant efficiency enhancement as
compared to the SQ limit in the case where the band is split,
because in such case a) the electron or hole decay rate through phonon
emission is reduced because of the mini-gaps, and b)
the impact ionization rate should be significantly increased
because the bands become flat\cite{Manousakis2019} and because of
reduced probability for carrier decay through phonon emission.
The impact ionization process leads to multiple carrier generation.

Namely, we find that the process of
decay via phonon emission  is halted by the presence of the minigaps in our proposed structures
until the excited electron finds a hole to recombine or excites another electron
from the valence band through the process of
impact ionization. Namely, the fragmented conduction and valence bands
work as a catalyst for impact ionization, because the small gaps block
the decay channel through phonons of the photoexcited electrons and holes.
\iffalse
The matrix element between the initial and final state, which enters in the
expression for the transition rate given by Fermi's golden rule when 
decay occurs,  is subject to the Wigner-Eckart
theorem which implies certain selection rules for allowed transitions
to occur. The incoming solar photon has definite energy and momentum and can
only cause a change in angular momentum of $\pm$ 1.
This means that for a photon to promote an electron from an occupied
state to an unoccupied state, the energy difference between the two
states has to be exactly the energy of the absorbed photon
and, in addition, the difference between the initial and final momentum of the
electron should be that carried by the photon. This can be easily satisfied when the
material contains a single band with states infinitesimally separated.
However, when
the band is split into small pieces, these conditions
and the additional condition that the transition has to be
from bands of different orbital character, i.e., made
from atomic orbitals differing in angular momentum by $\Delta L = \pm 1$,
are less likely to be simultaneously satisfied.
This leads to an increase to the impact ionization rate.
\fi

\begin{figure*}
  \begin{center}
            \subfigure[]{
              \includegraphics[scale=0.3]{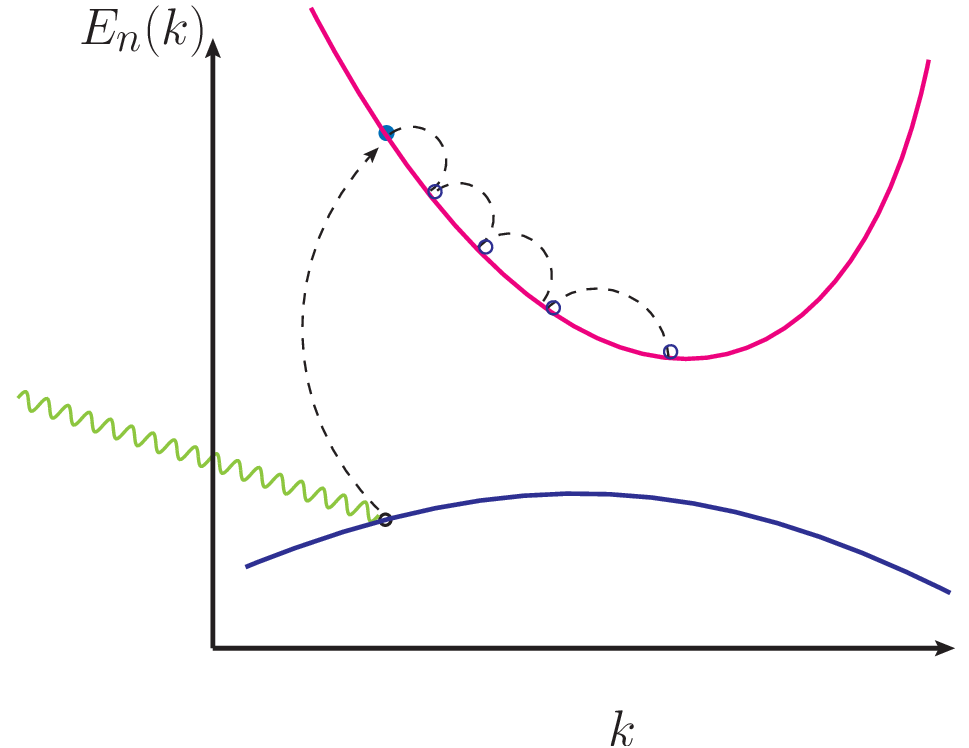} 
              }
            \subfigure[]{
              \includegraphics[scale=0.3]{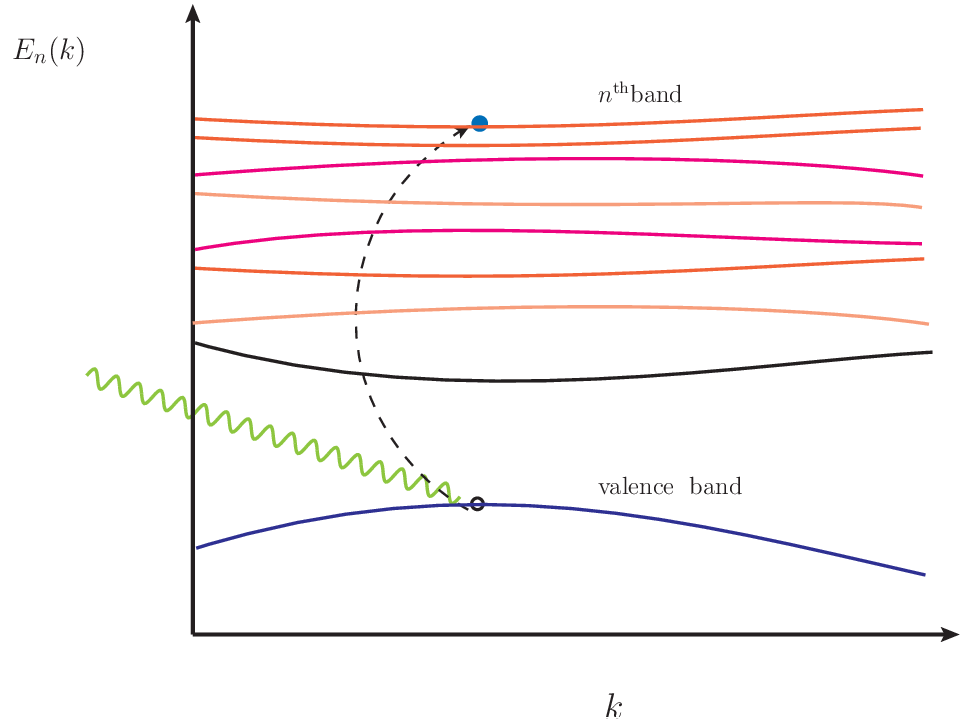} 
              }
            \subfigure[]{
              \includegraphics[scale=0.3]{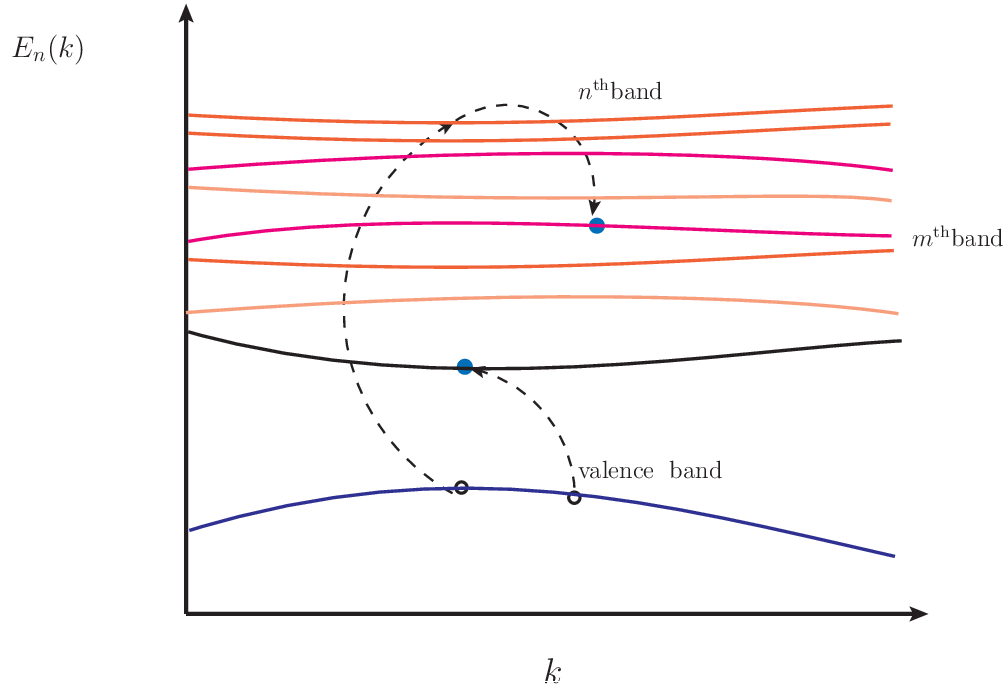}
              }
    \caption{See main text for explanation.}
  \label{fig5}
  \end{center}
\end{figure*}

In a nutshell the mechanism can be described in simple terms as follows:
First, in the absence of band fragmentation, when the band-structure is as the one illustrated in Fig.~\ref{fig5}(a),
the incident solar photon promotes an electron from the
      valence band to the conduction band significantly above the
      CBM. Subsequently, the electron
      decays to the CBM via a series phonon emission processes. 
In such case the excess energy, beyond the large energy gap, is consumed by phonons, which leads
to a significant limitation of the PV efficiency\cite{10.1063/1.1736034}.
 In the case of band fragmentation (Fig.~\ref{fig5}(b)), when the same incident solar photon promotes an electron to a
high-energy mini-band $n$ with energy $E_n({\bf k})$, and bandwidth $W_n$,
separated by the lower energy band $n-1$ by a gap $G_n$,
the excited electron can only lose a small
part of its energy by emitting phonons. The presence of the
small gap $G_n$, between successive bands, prevents further phonon
emission.
Furthermore, the mechanisms left for the electron to lose more energy are only
a) recombination with a hole or b) to fall into a lower energy band
and simultaneously promote another valence electron to a low energy band.
This second process (illustrated in Fig.~\ref{fig5}(c)) is the well-known
impact ionization process which leads to multi-carrier
generation\cite{PhysRevB.82.125109,PhysRevB.90.165142,Manousakis2019}.
The photon produced by process (a) of recombination can also generate a
different electron/hole pair, also leading to carrier multiplication,
thus, increasing efficiency further. We have not included this latter
efficiency increasing process\cite{Vos1980}. We have included, however,
in addition to the
impact ionization and Auger recombination processes, the role
of carrier recombination, and of
electron-electron, hole-hole and electron-hole scattering.

For completeness we would like to mention that for reasons somewhat related to what is proposed here,
sometime ago, it was suggested that  
using confining geometries, such as quantum wells,
quantum wires, quantum dots and nanostructures,
the relaxational time scales can be significantly 
affected \cite{10.1063/1.327889,PhysRevB.44.10945,PhysRevB.42.8947,PhysRevB.51.13281}
which would allow the possibility for impact ionization.

In conclusion, we have demonstrated that splitting the bands by means
of growing multilayers of alternating lattice-mismatched atomically-thin
layers of semiconductors or by means of alternating twisted multilayers of the
same atomically-thin layers, can lead to a significant
improvement in the efficiency of converting the broad solar spectrum
into electrical energy.

\section{Acknowledgments}

This work was supported by the U.S. National Science Foundation under Grant No. NSF-EPM-2110814.

\appendix

\section{Tight-binding treatment of the strained $3\times 3$ bilayer}
\label{2d-mismatch}

\begin{figure}[htb]
  \begin{center}
              \includegraphics[scale=0.4] {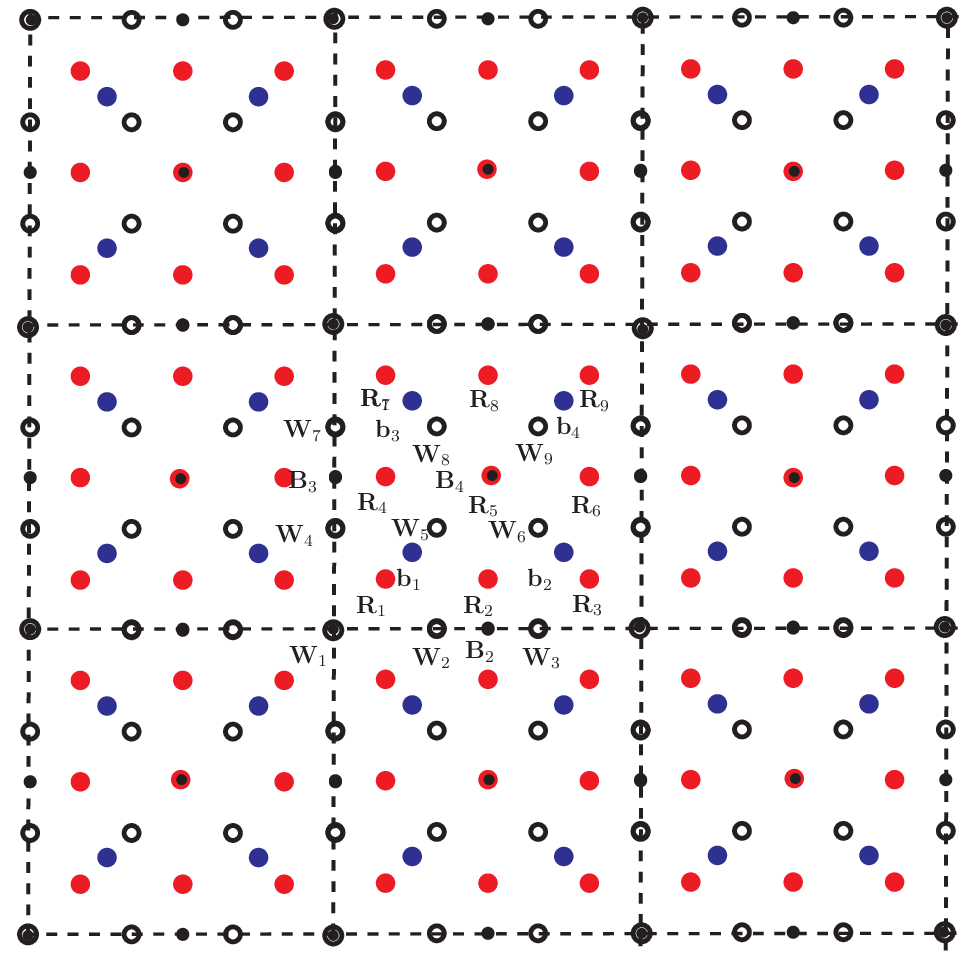}
                \caption{The same lattice as the one of Fig.~\ref{fig}(b) where
                  we have marked the atoms in order to define the matrix elements       of Table~\ref{table:1}}
  \label{fig10}
  \end{center}
\end{figure}

In this part of the appendix, we define the tight-binding matrix elements
of the case of the strained (mismatched) $3\times 3$ bilayer.
We will use Fig.~\ref{fig10} to define the various
hopping matrix elements of our tight-binding model.

For the bottom layer (layer 1) we use on-site energies for the black and blue sites:
  $E_B=0$ $ E_b = 1.0$ eV.
We only consider the hopping matrix element between black and its four diagonally neighboring blue sites and we take it to be 
$  t = 1 \mathrm{eV}$.

Similarly for the top layer (layer 2) there are white and red sites
with on-site energies: 
  $E_W=0$ $E_R = 1.0$ {eV}.
The white sites form a square lattice of lattice constant $a^{\prime}=2/3 a$ with a red site in the center. The only hopping matrix element in the second layer is between white and its four diagonally neighboring red sites and we take it as
$  t^{\prime} = 0.5$ {eV}.

We introduce the following interlayer hopping matrix elements.
When a white site is exactly on top of a black site there is a hopping
$  t_0 = 0.4$ {eV}.
When a red site is exactly on top of a black site there is a hopping
$  t^{\prime}_0 = 0.4~\mathrm{eV}$.
We allow a hopping matrix element of $  t_1 = 0.3~\mathrm{eV}$ between a black and the nearest white site (which is either along the
$x$ or $y$ direction relative to the black site, such as $W_2$ and $B_2$ sites
or $B_3$ and $W_7$ sites in Fig.~\ref{fig10}).
Similarly we allow a hopping $  t^{\prime}_1 = 0.3~\mathrm{eV}$
between a black and the nearest red site (which is either along the
$x$ or $y$ direction relative to the black site, such as $B_2$ and $R_2$ sites
or $B_3$ and $R_4$ cites in Fig.~\ref{fig10}).
Moreover, a hopping $  t_2 = 0.2~\mathrm{eV}$ is used
 between a black and the nearest red cite (which is
is at 45$^{\circ}$ relative to the black site, such as $B_1$ and $R_1$ sites
 in Fig.~\ref{fig10}).
A hopping  matrix element $ t^{\prime}_2 = 0.2~\mathrm{eV}$ is used
 between a black and the nearest white sites (which is
is at 45$^{\circ}$ relative to the black sites, such as $B_4$ and $W_5$ sites
 in Fig.~\ref{fig10}).
 A hopping $  t_3 = 0.3~\mathrm{eV}$ between a blue and the nearest white
 site (which is
is at 45$^{\circ}$ relative to the black site, such as $b_1$ and $W_5$ sites
or $b_3$ and $W_8$ sites in Fig.~\ref{fig10}) is used.
Lastly, we use a hopping  $ t_4 = 0.3~\mathrm{eV}$ between a blue and the nearest red site (which is is at 45$^{\circ}$ relative to the black site, such as $b_1$ and $R_1$ sites
or $b_3$ and $R_7$ sites in Fig.~\ref{fig10}).

These matrix elements leads to the tight-binding matrix
in momentum space given in Table~\ref{table:1}.
\begin{table*}
  \scriptsize
		\begin{tabular}{ |c|c|c|c|c|c|c|c|c|c|c|c|c|c|c|c|c|c|c|c|c|c|c|c|c|c|c|}
		    \hline
     & ${\bf B}_1$ & ${\bf B}_2$ & ${\bf B}_3$ & ${\bf B}_4$ & ${\bf W}_1$ &${\bf W}_2$ &${\bf W}_3$ &${\bf W}_4$ &${\bf W}_5$ &${\bf W}_6$ &${\bf W}_7$ &${\bf W}_8$ &${\bf W}_9$ & ${\bf R}_1$ &${\bf R}_2$ &${\bf R}_3$     &${\bf R}_4$ &${\bf R}_5$ &${\bf R}_6$ &   ${\bf R}_7$   &${\bf R}_8$ &${\bf R}_9$    & ${\bf b}_1$  &${\bf b}_2$    &${\bf b}_3$     & ${\bf b}_4$  \\
                    \hline
${\bf B}_1$ & $E_B$ &   0   & 0    &   0   & 0     & 0    & 0    & 0   & 0    & 0    & 0    & 0    & 0   & $t_1$  & 0   &$t^{-x}_1$ &   0  & 0    & 0    & $t^{-y}_1$&   t  & $t^{-x-y}$ & $ t$ & $t^{-x}$ & $t^{-y}$ & $t^{-x-y}$ \\
			\hline
${\bf B}_2$ & 0     &$E_B$  &   0  &  0    &   0   & 0    & $t_2$&$t_2$ & 0    & 0    & 0    & 0    & 0   & $t_1$  & 0   &$t^{-x}_1$ &   0  & 0    & 0    & $t^{-y}_1$&   t  & $t^{-x-y}$ & $ t$ & $t^{-x}$ & $t^{-y}$ & $t^{-x-y}$ \\
			\hline
${\bf B}_3$ & 0 & 0    &$E_B$   &  0    &   0   & 0    & 0 &$t_1$ & 0    & 0    & $t_1$    & 0    & 0   & 0  & 0   & 0  & $t^{\prime}_1$    & 0    & $t^{\prime -x}_1$&  0  & 0 & 0 & $t$ & $t^{-x}$ & $t$ & $t^{-x}$ \\
			\hline
${\bf B}_4$ & 0     &   0  &  0    & $E_B$  &   0   & 0   & 0 & 0  & $t_1$&$t_1$ & 0    & $t_1$  & $t_1$& 0 & 0 & 0 & 0  &$t^{\prime}_0$ &   0  & 0    & 0    & 0  & $t$ & $ t$ & $t$ & $t$ \\
			\hline
${\bf W}_1$ & $t_0$     &   0  &  0    & 0  &   $E_W$   & 0   & 0 & 0  & 0 & 0 & 0    & 0  & 0 & $t^{\prime}$ & 0 & $t^{\prime -x}$ & 0  & 0 &   0  & $t^{\prime -y}$    & 0    & $t^{\prime -x-y}$  & 0 & 0 & 0 & 0 \\
			\hline
${\bf W}_2$ & 0 & $t_1$   &  0    & 0  &  0 &  $E_W$    & 0 & 0  & 0 & 0 & 0    & 0  & 0 & 0 & $t^{\prime}$ & 0 & 0 &0 & 0 & $t^{\prime -y}$ & $t^{\prime -y}$    & 0  & 0 & 0 & 0 & 0 \\
			\hline
${\bf W}_3$ & 0 & $t_1$   &  0    & 0  &  0 & 0 &  $E_W$    & 0  & 0 & 0 & 0    & 0  & 0 & 0 & $t^{\prime}$ &$t^{\prime}$ & 0 & 0 &0 & 0 & $t^{\prime -y}$ & $t^{\prime -y}$ & 0 & 0 & 0 & 0 \\
			\hline
${\bf W}_4$ & 0 & 0 & $t_1$   &  0    & 0  &  0 & 0 &  $E_W$    & 0  & 0 & 0 & 0    & 0  & $t^{\prime}$ & 0 & $t^{\prime -x}$& $t^{\prime }$ & 0 & $t^{\prime -x}$ & 0 &0 & 0 & 0 & 0 & 0 & 0 \\
			\hline
${\bf W}_5$ & 0 & 0 & 0 & $t^{\prime}_2$   &  0    & 0  &  0 & 0 &  $E_W$    & 0  & 0 & 0 & 0   & $t^{\prime}$ & $t^{\prime }$& 0 & $t^{\prime }$ & $t^{\prime}$ & 0 &0 & 0 & 0 & $t_3$ & 0 & 0 & 0 \\
			\hline
${\bf W}_6$ & 0 & 0 & 0 & $t^{\prime}_2$   &  0    & 0  &  0 & 0 & 0 &  $E_W$    & 0  & 0 & 0 & 0   & $t^{\prime}$ & $t^{\prime }$& 0 & $t^{\prime }$ & $t^{\prime}$ & 0 &0 & 0 & 0 & $t_3$ & 0 & 0 \\
			\hline
${\bf W}_7$ & 0 & 0 & $t_1$   &  0    & 0  &  0 & 0 & 0 & 0 & 0& $E_W$    & 0  & 0 & 0 & 0 & 0  & $t^{\prime}$ & 0 &  $t^{\prime -x}$ & $t^{\prime}$ & 0 &  $t^{\prime -x}$ & 0 & 0 & 0 & 0 \\
			\hline
${\bf W}_8$ & 0 & 0 & 0 & $t^{\prime}_2$   &  0    & 0  &  0 & 0 & 0 & 0 & 0& $E_W$    & 0  & 0 & 0 & 0  & $t^{\prime}$ & $t^{\prime }$ &0 & $t^{\prime}$ & $t^{\prime}$ & 0 & 0 & 0 &$t_3$ & 0 \\
			\hline
${\bf W}_9$ & 0 & 0 & 0 & $t^{\prime}_2$   &  0    & 0  &  0 & 0 & 0 & 0 & 0 & 0& $E_W$    & 0  & 0 & 0 & 0  & $t^{\prime}$ & $t^{\prime }$ &0 & $t^{\prime}$ & $t^{\prime}$ & 0 & 0 & 0 &$t_3$ \\
			\hline
${\bf R}_1$ & $t_2$   &  0 & 0 & 0 & $t^{\prime}$ & $t^{\prime }$ &0 & $t^{\prime}$ & $t^{\prime}$ & 0    & 0  &  0 & 0 & $E_R$    & 0  & 0 & 0 & 0   & 0 & 0 & 0 &0 &$t_4$ & 0 & 0 & 0 \\
			\hline
${\bf R}_2$ & 0 & $t^{\prime}_1$   &  0 & 0 & 0 & $t^{\prime}$ & $t^{\prime }$ &0 & $t^{\prime}$ & $t^{\prime}$ & 0    & 0  &  0 & 0 & $E_R$    & 0  & 0 & 0 & 0   & 0 & 0 & 0 &0 & 0 & 0 & 0 \\
			\hline
$R_3$ & $t^{+x}_2$ & 0 &0 & 0 & $t^{\prime +x}$ & 0 &  $t^{\prime }$ & $t^{\prime +x}$ & 0 & $t^{\prime}$ & 0    & 0  &  0 & 0 & 0 & $E_R$    & 0  & 0 & 0 & 0   & 0 & 0 & 0  & $t_4$ & 0 & 0 \\
			\hline
${\bf R}_4$ &0   &  0 & $t^{\prime}_1$ & 0 &0 & 0 & 0 & $t^{\prime }$ & $t^{\prime }$ & 0 & $t^{\prime}$ & $t^{\prime}$ & 0    & 0  &  0 & 0 & $E_R$    & 0  & 0 & 0 & 0   & 0 & 0 & 0  & 0 & 0 \\
			\hline
${\bf R}_5$ &0   &  0  & 0 & $t^{\prime}_0$ & 0 &0 & 0 & 0 & $t^{\prime }$ & $t^{\prime }$ & 0 & $t^{\prime}$ & $t^{\prime}$ & 0    &  0 & 0 & 0&  $E_R$    & 0  & 0 & 0 & 0   & 0 & 0 & 0 & 0 \\
			\hline
${\bf R}_6$ &0    & 0 & $t^{\prime +x}_1$ & 0 &0 & 0 & 0 & $t^{\prime +x}$ & 0 & $t^{\prime }$ & $t^{\prime +x}$& 0 & $t^{\prime}$ & 0    &  0 & 0 & 0& 0 &  $E_R$    & 0 & 0 & 0   & 0 & 0 & 0 & 0 \\
			\hline
${\bf R}_7$ & $t^{+y}_2$ & 0 &0 & 0 & $t^{\prime +y}$ & $t^{\prime +y}$ & 0 & 0 & 0 & 0 & $t^{\prime}$ & $t^{\prime}$ & 0    &  0 & 0 & 0& 0 & 0& 0 & $E_R$    & 0 & 0 & 0   & 0 & $t_4$ & 0 \\
			\hline
${\bf R}_8$ & 0 & $t^{\prime +y}_1$ & 0 &0 & 0 & $t^{\prime +y}$ & $t^{\prime +y}$ & 0 & 0 & 0 & 0 & $t^{\prime}$ & $t^{\prime}$ & 0    &  0 & 0 & 0& 0 & 0& 0 & $E_R$    & 0 & 0 & 0   & 0  & 0 \\
			\hline
${\bf R}_9$ & $t^{+x+y}_2$ & 0 &0 & 0 & $t^{\prime x+y}$ & 0 & $t^{\prime +y}$ & 0 & 0 & 0 & $t^{\prime +x}$ & 0 & $t^{\prime}$ & 0    &  0 & 0 & 0& 0 & 0& 0 & 0 & $E_R$    & 0 & 0 &0 & $t_4$ \\
			\hline
${\bf b}_1$ & $t$ & $t$ & $t$ & $t$ & 0 &0 & 0 &0 & $t_3$ & 0 & 0 & 0 & 0 & $t_4$ & 0 & 0    &  0 & 0 & 0& 0 & 0& 0 & $E_b$    & 0 & 0 & 0 \\
			\hline
${\bf b}_2$ & $t^{+x}$ & $t$ & $t^{+x}$ & $t$ & 0 &0 & 0 & 0 &0 & $t_3$ & 0 & 0 & 0 & 0 & 0 &  $t_4$ & 0 & 0    &  0 & 0 & 0& 0 & 0& $E_b$   & 0 & 0 \\
			\hline
${\bf b}_3$ & $t^{+y}$ & $t^{+y}$ & $t$ & $t$ & 0 &0 & 0 & 0 & 0 &0 & 0 & $t_3$ & 0 & 0 & 0 & 0 & 0 & 0 & 0 & $t_4$ & 0 & 0    &  0 & 0 & $E_b$   &  0 \\
			\hline
${\bf b}_3$ & $t^{x+y}$ & $t^{+y}$ & $t^{+x}$ & $t$ & 0 &0 & 0 & 0 & 0 &0 & 0 & 0 & $t_3$ & 0 & 0 & 0 & 0 & 0 & 0 & 0 & 0 & $t_4$ & 0 & 0 & 0 & $E_b$  \\
			\hline
		\end{tabular}
		\caption{ }
		\label{table:1}
                \vskip 0.4 in
\end{table*}

Our results of the diagonalization of this matrix for the parameter values
discussed above leads to the band-structure
discussed in the main part of the paper.

\section{Brillouin zone of twisted bilayer}
\label{twisted-bz}

\begin{figure*}
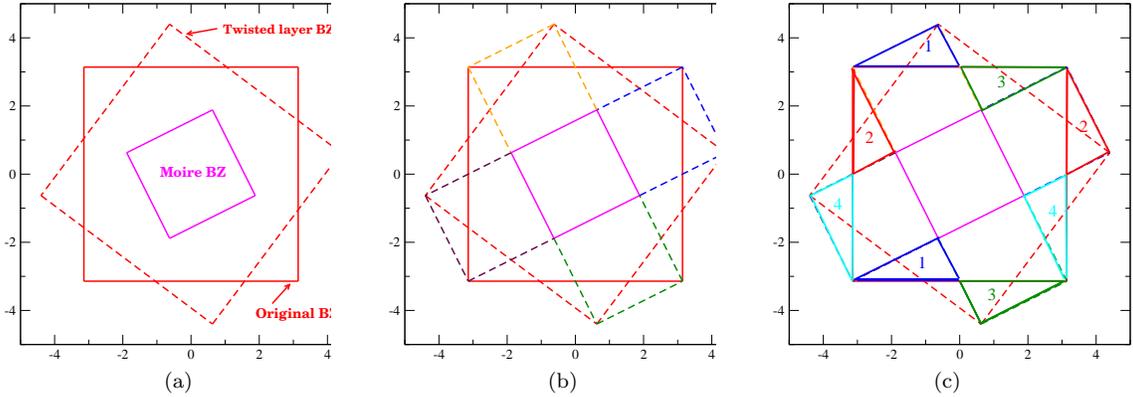

  \begin{center}
            \subfigure[]{
              \includegraphics[scale=0.3]{Fig14a.eps} 
              }
            \subfigure[]{
              \includegraphics[scale=0.3] {Fig14b.eps}
            }
            \subfigure[]{
              \includegraphics[scale=0.3] {Fig14c.eps}
            }
    \caption{(a) The large-size Brillouin
      zones shown are those of each of the layers, while the smaller-size
    Brillouin zone corresponds to the combined moir\'e-structure. (b) The moir\'e BZ is expanded in the 4 directions using 4 reciprocal lattice vectors. (c) Demonstration that the BZ moir\'e BZ and it 4 extensions cover completely the large BZ of the bottom layer by cutting and pasting
    the pieces 1,2,3,4.}
  \label{fig11}
  \end{center}
\end{figure*}

The moir\'e BZ is shown in the right panel of Fig.~\ref{fig11}, along with those
of each individual layer.
When we fold the 2 bands of each layer inside the moir\'e BZ, as schematically illustrated in Fig.~\ref{fig11}, we obtain the non-interacting bands shown by broken magenta lines in Fig.~\ref{fig4}(b) of the main manuscript.  Fig.~\ref{fig11}(b) shows the moir\'e BZ is expanded in the 4 directions using 4 reciprocal lattice vectors. Fig.~\ref{fig11}(c) demonstrates that the BZ moir\'e BZ and it 4 extensions cover completely the large BZ of the bottom layer by cutting and pasting
    the pieces 1,2,3,4.

\end{document}